\documentclass[journal,twoside,web]{ieeecolor}
\usepackage{generic}
\usepackage{cite}
\usepackage{amsmath,amssymb,amsfonts}
\usepackage{algorithmic}
\usepackage{graphicx}
\usepackage{algorithm,algorithmic}
\usepackage{amssymb}
\usepackage{hyperref}
\hypersetup{hidelinks=true}
\usepackage{textcomp}
\usepackage{wrapfig}
\usepackage{nicematrix}

\usepackage{tikz}
\usepackage{tikz-cd} 
\usepackage{caption}  
\usetikzlibrary{matrix}
\usepackage{graphicx}
\usetikzlibrary{matrix,arrows,decorations.pathmorphing}
\usepackage{arydshln}

\newtheorem{cor}{Corollary}
\newtheorem{definition}{Definition}

\newtheorem{lemma}{Lemma}
\newtheorem{remark}{Remark}
\newtheorem{thm}{Theorem}

\DeclareMathOperator{\rank}{rank}
\DeclareMathOperator{\R}{\mathbb{R}}
\DeclareMathOperator{\C}{\mathbb{C}}

\def\BibTeX{{\rm B\kern-.05em{\sc i\kern-.025em b}\kern-.08em
    T\kern-.1667em\lower.7ex\hbox{E}\kern-.125emX}}
\begin{document}
\title{Functional Controllability, Functional Stabilizability, and the Generalized Separation Principle}
\author{Tyrone Fernando and Mohamed Darouach
	%\IEEEmembership{Senior Member, IEEE}
	\thanks {This work was supported by Gledden Senior Visiting Fellowship, Institute of Advanced Studies (IAS), UWA, Australia. (Corresponding Author: Tyrone Fernando.)}
	\thanks{T. Fernando is with the Department of Electrical, Electronic and Computer Engineering, University of Western Australia (UWA), 35 Stirling Highway, Crawley, WA 6009, Australia. (email:  tyrone.fernando@uwa.edu.au)}
	\thanks{M. Darouach is with the University de Lorraine, Centre de Recherche en Automatique de Nancy (CRAN UMR-7039, CNRS), IUT de Longwy, 186 rue de Lorraine 54400, Cosnes et Romain, France. (email:  mohamed.darouach@univ-lorraine.fr)}
}

\maketitle

\begin{abstract}
	This paper introduces the new concepts of Functional Controllability and Functional Stabilizability, and establishes their duality with Functional Observability and Functional Detectability, respectively. A Generalized Separation Principle is presented, under which the classical Separation Principle emerges as a special case. Conditions for the existence of functional controllers of a specified order are derived. Notably, the proposed design framework does not require full controllability. In addition, a functional observer-based controller design is developed for systems that may be both uncontrollable and unobservable. The results presented extend and generalize the classical full-state observer based feedback control paradigm.
\end{abstract}


\section{Introduction} 

The concepts of Controllability and Observability~\cite{ref1}, \cite{ref2}, originally introduced by Kalman, are foundational to modern control theory and have shaped the development of analysis and design methodologies for dynamical systems. These properties underpin the classical separation principle~\cite{ref2}, \cite{ref3}, which enables the independent design of observers and controllers in observer-based feedback configurations. In this framework, controllability ensures that the entire state vector can be driven to any desired value via suitable control inputs, while observability guarantees that the full state can be reconstructed from output measurements over time. The separation principle asserts that, under both controllability and observability, one can design the observer and controller independently without compromising closed-loop stability.

While powerful, this framework assumes full-state controllability and observability—assumptions that may be unnecessarily restrictive or impractical in many modern applications. In various real-world systems—such as energy networks~\cite{ref4}, epidemic spread models~\cite{ref5}, and multi-robot systems~\cite{ref6}—the objective is often to control or estimate only specific quantities of interest, typically expressible as linear functionals of the state. For example, one may aim to regulate aggregate active power or battery state-of-charge in energy systems, or control the position of an end-effector in robotic applications. Enforcing full-state estimation or control in such settings can lead to over-engineered solutions or exclude viable designs altogether.
This motivates a shift to a more targeted, function-oriented framework focused on estimating and controlling relevant state functionals. Such an approach is particularly valuable in large-scale or networked systems, or in applications involving privacy, abstraction, or dimensionality reduction~\cite{16new}--\cite{15new}. To formalise this idea, the notions of \emph{functional observability} and \emph{functional detectability} were introduced in~\cite{ref10,ref12}, and have since generated considerable interest (see, e.g.,~\cite{ref5,ref11,20a,ref13,16new,17new}). Developments on the design of functional observers can be found in~\cite{ref13}--\cite{3}. 

In this paper, the dual counterparts—\emph{functional controllability} and \emph{functional stabilizability}—are introduced. These properties generalize their classical analogues by focusing on specific linear combinations of the state vector rather than the full state. Functional controllability refers to the ability to steer a particular state functional using admissible inputs, while functional observability pertains to the ability to reconstruct a functional of the state from output measurements. Crucially, these generalizations retain the mathematical structure and rigor of classical control theory while offering improved flexibility and applicability.
A concept closely related to functional controllability is \emph{target output controllability} (also known as output controllability)~\cite{1a}--\cite{3a}. The key distinction lies in the control objective: target output controllability seeks to steer the output \( z(t) \) from an initial state \( z(t_0) \) to a desired final value \( z(t_1) \) in finite time. In contrast, functional controllability allows for asymptotic convergence to a target value \( z^* \), i.e.,
\[
z^* = \lim_{t \to \infty} z(t)
\]
with an arbitrarily specified convergence rate. As will be shown in the sequel, when \( z(t) = x(t) \), both notions reduce to classical full-state controllability. For this reason, target output controllability has often been regarded as the generalization of classical controllability. However, this paper argues that \emph{functional controllability} offers a more natural and structurally consistent generalization. 
	Notably, a recent study by Montanari et al.~\cite{3a} demonstrated that the duality between target output controllability and functional observability holds only under specific conditions, distinguishing between strong and weak forms of duality. In contrast, the results presented here establish that functional controllability and functional observability are inherently dual concepts.

The main contributions of this paper are as follows:
\begin{itemize}
	\item[(i)] The concepts of \emph{functional controllability} and \emph{functional stabilizability} are formally defined, and their duality with \emph{functional observability} and \emph{functional detectability} is established.
	\item[(ii)] A \emph{generalized separation principle} is developed, enabling the independent design of functional controllers and observers.
	\item[(iii)] Necessary and sufficient conditions are derived for the existence of functional controllers of a specified order.
	\item[(iv)] It is shown that the classical separation principle emerges as a special case of the generalized framework.
\end{itemize}

By relaxing the requirement for full-state controllability and observability, the presented results significantly broaden the scope of observer-based control design and provide a principled foundation for more efficient and targeted control system synthesis.

\section{Problem Formulation, Notation and Preliminaries}
Consider a continuous-time, linear time-invariant system described by the triple \( (A, B, C) \) of the following form:
\begin{IEEEeqnarray}{rcl}\label{Rotella}
	\dot{x}(t) &=& Ax(t)+Bu(t) \IEEEyessubnumber \label{1a}\\
	y(t) &=& Cx(t) \IEEEyessubnumber \label{1b} 
\end{IEEEeqnarray}
with $x(t)\in \R^n$ is the state  vector, $u(t)\in \R^m$ is the known input, and $y(t)\in \R^p$ is the measurement output. Matrix $A\in \R^{n\times n}$, matrix $B\in \R^{n\times m}$ and the full row rank matrix $C\in \R^{p\times n}$ define the system dynamics, input matrix, and output matrix, respectively.

In contrast to classical control design, which aims to estimate or control the entire state vector \( x(t) \), the focus here is on estimating and controlling specific linear functionals of the state. Let \( F \in \mathbb{R}^{r \times n} \) define the desired functional output \( z(t) \in \mathbb{R}^r \):
\begin{equation}
	z(t) = Fx(t) \nonumber
\end{equation}
where $F$ is a full row rank matrix. Matrix $F$ is taken as a full row rank matrix because the linear functions to be controlled are all linearly independent, ensuring that any linearly dependent functions are also controlled. \\

\noindent The objective is to design:
\begin{itemize}
	\item A control signal $u(t)$ that drives $z(t)$ to zero (or to a desired trajectory) at a specified rate, and
	\item An observer to estimate 
	$z(t)$ using only the available output $y(t)$,
\end{itemize}
without requiring full controllability or observability of the original system \eqref{1a}-\eqref{1b}.\\

\noindent This leads to several key questions:
\begin{itemize}
	\item Under what conditions can a given functional $z(t)=Fx(t)$ be controlled or stabilised independently of the entire state?
	\item Under what conditions can $z(t)$ be estimated from the output 
	$y(t)$, even when the full state is not observable?
	\item Can controller and observer design be decoupled, as in the classical separation principle, in this functional setting?
\end{itemize}

To rigorously investigate the questions above, the notions of functional controllability and functional stabilizability are formalised within this framework, and conditions are derived under which independent controller and observer design is possible. This development also clarifies how functional controllability differs from related concepts such as target output controllability.
The paper is structured as follows. Section III presents the criteria for target output controllability, functional controllability, and functional stabilizability, and establishes the dual relationships between these concepts and their observable counterparts. Section~IV derives the existence conditions and design procedure for functional controllers via the placement of \( r \) closed-loop poles, where \( r \) corresponds to the number of functionals to be controlled.
 Section V introduces and formalises the generalized separation principle, extending the classical result to functional observer-based control. Finally, Sections VI and VII provide numerical examples and concluding remarks, respectively.

Throughout the paper, \( G^T \) denotes the transpose of a matrix \( G \), and \( G^- \) a generalized inverse satisfying \( GG^-G = G \). The rank, image, and kernel of \( G \) are denoted by \( \mathrm{rank}(G) \), \( \mathrm{Im}(G) \), and \( \ker(G) \), respectively. The symbol \( \oplus \) denotes the direct sum of subspaces. For a matrix \( G \in \mathbb{R}^{r \times n} \), \( G^\perp \in \mathbb{R}^{(n - r) \times n} \) denotes a matrix whose rows form a basis for the orthogonal complement of the row space of \( G \), i.e., \( G^\perp G^T = 0 \). Identity matrices are denoted by \( I \), with subscripts when necessary to indicate dimension and \( \mathfrak{Re}(\lambda) \) denotes the real part of a complex scalar \( \lambda \).
Unstable modes are defined as the eigenvectors (or generalized eigenvectors) of \( A \) associated with eigenvalues having non-negative real parts, i.e., \( \mathfrak{Re}(\lambda) \geq 0 \).
For appropriate dimensions of \( A \in \mathbb{R}^{n \times n} \) and \( G \in \mathbb{R}^{n \times k} \) or \( G \in \mathbb{R}^{k \times n} \), $k\in\{1,\dots,n\}$, the standard controllability and observability matrices associated with the pair \( (A, G) \) are defined, respectively, as
\[
\mathcal{C}_{(A,G)} = \begin{pmatrix}
	G & AG & \cdots & A^{n-1}G
\end{pmatrix}, \,\,
\mathcal{O}_{(A,G)} = \begin{pmatrix}
	G \\ GA \\ \vdots \\ GA^{n-1}
\end{pmatrix}
\]
where \( G \) is interpreted as an input matrix, i.e.,  \( G \in \mathbb{R}^{n \times k} \) when constructing \( \mathcal{C}_{(A,G)} \), and as an output matrix, i.e., \( G \in \mathbb{R}^{k \times n} \) when constructing \( \mathcal{O}_{(A,G)} \).
Accordingly, the controllable and uncontrollable subspaces of the pair \( (A, G) \) are defined as \( \mathrm{Im}(\mathcal{C}_{(A,G)}) \) and \( \ker(\mathcal{C}_{(A,G)}^T) \), respectively, when \( G \) is interpreted as an input matrix. Similarly, the observable and unobservable subspaces are \( \mathrm{Im}(\mathcal{O}_{(A,G)}^T) \) and \( \ker(\mathcal{O}_{(A,G)}) \), respectively, when \( G \) is interpreted as an output matrix. Moreover, for $M \in \R^{k\times n}$, $k\in\{1,\dots,n\}$ the state space \( \mathbb{R}^n \) admits the orthogonal decompositions
\[
\mathbb{R}^n =  \mathrm{Im}(\mathcal{C}_{(A,M^T)})  \oplus  \ker(\mathcal{C}_{(A,M^T)}^T)
= \mathrm{Im}(\mathcal{O}_{(A,M)}^T)  \oplus \ker(\mathcal{O}_{(A,M)}).
\]
	Finally, note that the dual of a system described by the triple \( (A, B, C) \) is given by \( (A^T, C^T, B^T) \), under which controllability and observability exchange roles. This classical duality underpins the functional controllability–observability and functional stabilizability-detectability relationships established in this work.

\section{Criteria for Target Output Controllability, Functional Stabilizability and Functional Controllability}
The well-known concept of target output controllability and the introduced concepts of functional controllability and functional stabilizability in this paper are closely linked, yet each possesses its own subtle distinctions. To explore these nuances, the following definitions are first introduced:
\begin{definition}
	The linear combinations of states $z(t)=Fx(t)$ of system \eqref{Rotella}, or the triple $(A,B,F)$ is target output controllable, if for any initial state $z(t_0)$ and any final target state $z(t_1)$, there exists an input $u(t)$ that steers $z(t_0) = F x(t_0)$ to $z(t_1) = F x(t_1)$ in time $t_1 > t_0$.
\end{definition}

\begin{definition}
	The functional \( z(t) = Fx(t) \) is said to be controllable, or the triple \( (A, B, F^T) \) is said to be \emph{functional controllable}, if and only if every vector in the uncontrollable subspace of the pair \( (A, B) \) also lies in the uncontrollable subspace of the pair \( (A, F^T) \). That is,
	\[
	\ker\left(\mathcal{C}_{(A,B)}^T\right) \subseteq \ker\left(\mathcal{C}_{(A,F^T)}^T\right).
	\]
\end{definition}

\begin{definition}[Algebraic form of Definition 2]
	The functional \( z(t) = Fx(t) \) is controllable, or the triple \( (A, B, F^T) \) is said to be \emph{functional controllable}, if and only if for all \( v \in \mathbb{C}^n \)
	\[
	v^T\mathcal{C}_{(A, B)} = \mathbf{0} \quad \Rightarrow \quad v^T\mathcal{C}_{(A, F^T)}  = \mathbf{0}
	\]
	where \( \mathcal{C}_{(A, B)} \) and \( \mathcal{C}_{(A, F^T)} \) denote the controllability matrices corresponding to the input matrix \( B \) and matrix \( F^T \), respectively.
\end{definition}

\begin{definition}
	The functional \( z(t) = Fx(t) \) is stabilizable, or the triple \( (A, B, F^T) \) is \emph{functional stabilizable}, if and only if every left generalized eigenvector of \( A \) associated with an eigenvalue \( \lambda \in \mathbb{C} \)  satisfying \( \mathfrak{Re}(\lambda) \geq 0 \), which lies in the uncontrollable subspace corresponding to the pair \( (A, B) \), also lies in the uncontrollable subspace corresponding to the pair \( (A, F^T) \).
\end{definition}

\begin{definition}[Algebraic form of Definition 4]
	The functional \( z(t) = Fx(t) \) is stabilizable, or the triple \( (A, B, F^T) \) is  \emph{functional stabilizable}, if and only if for any \( \lambda \in \mathbb{C} \) with \( \mathfrak{Re}(\lambda) \geq 0 \) and for all \( v \in \mathbb{C}^n\)
	\begin{IEEEeqnarray}{rcl}
		&&v^T(\lambda I - A)^n  = \mathbf{0} \quad \text{and} \quad v^T\mathcal{C}_{(A, B)}  = \mathbf{0} \quad  \nonumber\\ 
		&&\Rightarrow  v^T\mathcal{C}_{(A, F^T)}  = \mathbf{0}. \nonumber
	\end{IEEEeqnarray}	
\end{definition}

The following theorem characterises target output controllability.

\begin{thm}[\cite{1b}]\label{thm1MD}
	The following conditions are equivalent:
	\begin{enumerate}
		\item The linear combination of states $z(t)=Fx(t)$ of system \eqref{Rotella} is target output controllable or the triple $(A,B,F)$ is target output controllable.
		\item $\rank(F\mathcal{C}_{(A,B)})=\rank(F)$, see \cite{1a}. 
		\item $\rank \begin{pmatrix}FB &F(A-\lambda I)B &\dots &F(A-\lambda I)^{n-1}B\end{pmatrix}$\\$=\rank(F), \forall \lambda \in \C$.
	\end{enumerate}
\end{thm}

The following theorem characterises functional controllability.

\begin{thm}\label{thm1MD1}
	The following conditions are equivalent:
	\begin{enumerate}
		\item The functional $z(t)=Fx(t)$ is controllable or the triple $(A,B,F^T)$ is functional controllable.
		\item $\rank\begin{pmatrix}\mathcal{C}_{(A, B)} &\mathcal{C}_{(A, F^T)}\end{pmatrix} = \rank\big(\begin{array}{ll}\mathcal{C}_{(A, B)}\end{array}\big)$
		\item $\rank\begin{pmatrix}\mathcal{C}_{(A, B)} & F^T\end{pmatrix} = \rank\big(\begin{array}{ll}\mathcal{C}_{(A, B)}\end{array}\big)$
	\end{enumerate}
\end{thm}

\begin{proof}
	From Definition 2 of functional controllability, the triple \( (A, B, F^T) \) is functional controllable if and only if
		\[
		\ker\left( \mathcal{C}_{(A, B)}^T \right) \subseteq \ker\left( \mathcal{C}_{(A, F^T)}^T \right).
		\]
	This can be restated as
	\[
	\ker\left( \mathcal{C}_{(A, B)}^T \right) \subseteq \ker\begin{pmatrix} \mathcal{C}_{(A, B)}^T \\ \mathcal{C}_{(A, F^T)}^T \end{pmatrix}.
	\]
	Since appending more rows (as in appending \( \mathcal{C}_{(A,F^T)}^T \)) can only further constrain the kernel, the following inclusions also hold:
	\[
	\ker \begin{pmatrix} \mathcal{C}_{(A, B)}^T \\ \mathcal{C}_{(A, F^T)}^T \end{pmatrix}  \subseteq \ker \begin{pmatrix} \mathcal{C}_{(A, B)}^T \\ F \end{pmatrix}  \subseteq \ker\left( \mathcal{C}_{(A, B)}^T \right)
	\]
	Therefore, all three kernel spaces are equal 
	\[
	\ker \begin{pmatrix} \mathcal{C}_{(A, B)}^T \\ \mathcal{C}_{(A, F^T)}^T \end{pmatrix}  = \ker \begin{pmatrix} \mathcal{C}_{(A, B)}^T \\ F \end{pmatrix}  = \ker\left( \mathcal{C}_{(A, B)}^T \right).
	\]
	Given that the matrices have the same number of columns and equal kernels, the rank--nullity theorem guarantees that their ranks are equal, and vice versa:
	\begin{IEEEeqnarray}{rcl}
		\rank \begin{pmatrix} \mathcal{C}_{(A, B)} & \mathcal{C}_{(A, F^T)} \end{pmatrix}  &=& \rank \begin{pmatrix} \mathcal{C}_{(A, B)} & F^T \end{pmatrix}  \nonumber\\
		&=& \rank\left( \mathcal{C}_{(A, B)} \right) \nonumber
	\end{IEEEeqnarray}
	which establishes the equivalence of all three items.
\end{proof}
The following theorem characterises functional stabilizability.
\begin{thm}
	The following conditions are equivalent:
	\begin{enumerate}
		\item The functional $z(t)=Fx(t)$ is stabilizable or the triple $(A,B,F^T)$ is functional stabilizable.
		\item $\rank 	\begin{pmatrix}
			(\lambda I -A)^n &\mathcal{C}_{(A,B)} &\mathcal{C}_{(A,F^T)}
		\end{pmatrix}$\\
		$=\rank 	\begin{pmatrix}
			(\lambda I -A)^n &
			\mathcal{C}_{(A,B)}
		\end{pmatrix}, \forall \lambda \in \C, \mathfrak{Re}(\lambda) \geq 0.
		$
		\item $\rank\left(\begin{array}{lll}(\lambda I -A)^n &\mathcal{C}_{(A, B)} &F^T\end{array}\right)$\\$=\rank\left(\begin{array}{ll}(\lambda I -A)^n &\mathcal{C}_{(A, B)}\end{array}\right), \forall \lambda \in \C,  \mathfrak{Re}(\lambda)\geq 0$.
	\end{enumerate}
\end{thm}

\begin{proof}
	From Definition 5, the triple \( (A, B, F^T) \) is functional stabilizable if and only if for each \( \lambda \in \mathbb{C} \) with \( \mathfrak{Re}(\lambda) \geq 0 \)
		\[
		\ker \begin{pmatrix} (\lambda I - A^T)^n \\ \mathcal{C}_{(A,B)}^T \end{pmatrix}
		\subseteq
		\ker \begin{pmatrix} (\lambda I - A^T)^n \\  \mathcal{C}_{(A,F^T)}^T \end{pmatrix}.
		\]
		This can be restated as
	\[
	\ker \begin{pmatrix} (\lambda I - A^T)^n \\ \mathcal{C}_{(A,B)}^T \end{pmatrix}
	\subseteq
	\ker \begin{pmatrix} (\lambda I - A^T)^n \\ \mathcal{C}_{(A,B)}^T \\ \mathcal{C}_{(A,F^T)}^T \end{pmatrix}.
	\]
Since appending more rows can only restrict the kernel,  the following inclusions also hold:
	\begin{IEEEeqnarray}{rcl}
		\ker \begin{pmatrix} (\lambda I - A^T)^n \\ \mathcal{C}_{(A,B)}^T \\ \mathcal{C}_{(A,F^T)}^T \end{pmatrix}
		&&\subseteq
		\ker \begin{pmatrix} (\lambda I - A^T)^n \\ \mathcal{C}_{(A,B)}^T \\ F \end{pmatrix} \nonumber\\
		&&\subseteq
		\ker \begin{pmatrix} (\lambda I - A^T)^n \\ \mathcal{C}_{(A,B)}^T \end{pmatrix}. \nonumber 
	\end{IEEEeqnarray}
	Hence, all three kernel spaces are equal
	\begin{IEEEeqnarray}{rcl}
		&&\ker \begin{pmatrix} (\lambda I - A^T)^n \\ \mathcal{C}_{(A,B)}^T \\ \mathcal{C}_{(A,F^T)}^T \end{pmatrix}
		=
		\ker \begin{pmatrix} (\lambda I - A^T)^n \\ \mathcal{C}_{(A,B)}^T \\ F \end{pmatrix} \nonumber\\
		&&=
		\ker \begin{pmatrix} (\lambda I - A^T)^n \\ \mathcal{C}_{(A,B)}^T \end{pmatrix}. \nonumber 
	\end{IEEEeqnarray}
Given that the matrices have the same number of columns and equal kernels, the rank--nullity theorem guarantees that their ranks are equal, and vice versa:
	\begin{IEEEeqnarray}{rcl}
		&&\rank \begin{pmatrix} (\lambda I - A)^n & \mathcal{C}_{(A,B)} & \mathcal{C}_{(A,F^T)} \end{pmatrix} \nonumber\\
		&&=
		\rank \begin{pmatrix} (\lambda I - A)^n & \mathcal{C}_{(A,B)} & F^T \end{pmatrix}\nonumber\\
		&&=
		\rank \begin{pmatrix} (\lambda I - A)^n & \mathcal{C}_{(A,B)} \end{pmatrix}.\nonumber
	\end{IEEEeqnarray}
	This establishes the equivalence of all three conditions.
\end{proof}

The following corollaries follow directly from Theorems 1, 2, and 3.

\begin{cor}
	If $F=I_n$, then the target output controllability condition 2) and also condition 3) of Theorem 1 reduce to the full state controllability condition of $\rank (\mathcal{C}_{(A,B)})=n$.
\end{cor}

\begin{cor}
	If $F=I_n$, then the functional controllability condition 2) and also condition 3) of Theorem 2 reduce to the full state controllability condition of $\rank (\mathcal{C}_{(A,B)})=n$.
\end{cor}

\begin{cor}
	If \( F = I_n \), then functional stabilizability conditions 2) and 3) of Theorem~3 reduce to the classical PBH test for stabilizability, i.e., for all \( \lambda \in \mathbb{C} \) with \( \mathfrak{Re}(\lambda) \geq 0 \),
	\begin{IEEEeqnarray}{rcl}
		\rank \begin{pmatrix}
			(\lambda I - A)^n & \mathcal{C}_{(A,B)}
		\end{pmatrix} 
		&=&
		\rank \begin{pmatrix}
			\lambda I - A & B
		\end{pmatrix}
		= n. \nonumber
	\end{IEEEeqnarray}
	This equivalence holds because
	\[
	\rank \begin{pmatrix}
		(\lambda I - A)^n & \mathcal{C}_{(A,B)}
	\end{pmatrix} = n 
	\;\Leftrightarrow\; 
	\ker \begin{pmatrix} (\lambda I - A^T)^n \\ \mathcal{C}_{(A,B)}^T \end{pmatrix} = \{ \mathbf{0} \}
	\]
	and for all \( \lambda \in \mathbb{C} \) with \( \mathfrak{Re}(\lambda) \geq 0 \), this kernel condition is equivalent to requiring that no generalized left eigenvector of \( A \) associated with eigenvalues in the closed right-half plane lies in the uncontrollable subspace of \( (A, B) \). That is, the pair \( (A, B) \) is stabilizable.
\end{cor}

A numerical example is presented to illustrate the subtle distinctions among the three concepts before delving into further details.

{\it Example 1:} Consider the following system with a state vector $x(t)=\begin{pmatrix}
	x_1(t)\\
	x_2(t)\\
	x_3(t)\\
	x_4(t)
\end{pmatrix}$ and matrices $A$ and $B$ as follows:
$$A=\begin{pmatrix}
	1 &0 &0&0\\
	0 &2&0&0\\
	0 &0 &-1&0\\
	0 &0 &0&3
\end{pmatrix} \mathrm{\,\,and\,\,} B=\begin{pmatrix}
	1 &0\\
	0 &1\\
	0 &0\\
	0 &0
\end{pmatrix}.$$
Consider four different functions to be controlled, defined as \( F = F_1, F_2, F_3, \) and \( F_4 \), where
\[
F_1^T=\begin{pmatrix}
	1\\
	1\\
	1\\
	1
\end{pmatrix},\quad
F_2^T=\begin{pmatrix}
	1\\
	1\\
	1\\
	0
\end{pmatrix},\quad
F_3^T=\begin{pmatrix}
	1\\
	1\\
	0\\
	0
\end{pmatrix},\quad
F_4^T=\begin{pmatrix}
	0\\
	0\\
	1\\
	1
\end{pmatrix}.
\]
Note that \( x_1(t) \) and \( x_2(t) \) are controllable, \( x_3(t) \) is stabilizable, and \( x_4(t) \) is uncontrollable. Based on Theorems~1, 2, and 3, the target output controllability, functional stabilizability, and functional controllability properties of the functionals \( z_1(t) = F_1 x(t) \), \( z_2(t) = F_2 x(t) \), \( z_3(t) = F_3 x(t) \), and \( z_4(t) = F_4 x(t) \) can be directly determined.
These outcomes are then related to the controllability, stabilizability, and uncontrollability properties of the individual state components of the system in Example~1. The relationships are summarised in Table~1.

\begin{table}[h!]
	\centering
	\renewcommand{\arraystretch}{1.5} % Adjust row height for readability
	% Define fixed column widths using p{}
	\begin{tabular}{@{}p{.8cm}|p{.7cm}p{.7cm}p{.7cm}p{.7cm}|p{.7cm}p{.7cm}p{.7cm}@{}}
		\hline
		\textbf{Func-} & \textbf{\(x_1(t)\)} & \textbf{\(x_2(t)\)} & \textbf{\(x_3(t)\)} & \textbf{\(x_4(t)\)} & \textbf{Target} & \textbf{Func.} & \textbf{Func.} \\ 
		\textbf{-tional} & \textbf{ctrb.} & \textbf{ctrb.} & \textbf{stbl.} & \textbf{unctrb.} & \textbf{outp.} & \textbf{stbl.} & \textbf{ctrb.} \\ 
		&  &  &  &  & \textbf{ctrb.} &  &  \\ 
		\hline
		\(z_1(t)\) & 1 & 1 & 1 & 1 & 1 & 0 & 0 \\ 
		\(z_2(t)\) & 1 & 1 & 1 & 0 & 1 & 1 & 0 \\ 
		\(z_3(t)\) & 1 & 1 & 0 & 0 & 1 & 1 & 1 \\ 
		\(z_4(t)\) & 0 & 0 & 1 & 1 & 0 & 0 & 0 \\ 
		\hline
	\end{tabular} 
	\caption{Summary of target output controllability, functional stabilizability, and functional controllability for the functionals \( z_i(t) = F_i x(t), i\in\{1,\dots,4\} \), and their correspondence with the controllable, stabilizable, and uncontrollable states of the system in Example~1.}
	\label{tab:compact_table}
\end{table}

Table~1 uses the following abbreviations: {\it ctrb} for controllable, {\it stbl} for stabilizable, {\it unctrb} for uncontrollable, {\it outp} for output, and {\it func} for functional. The digit 1 indicates "yes" and 0 indicates "no." The following observations can be made:

\begin{itemize}
	\item The functional $z_1(t)$ which is a mix of controllable, stabilizable and uncontrollable states (see the row corresponding to $z_1(t)$ in Table 1) is target output controllable whilst it is not functional stabilizable and also not functional controllable. 
	\item The functional $z_2(t)$ which is a mix of controllable and stabilizable states  (see the row corresponding to $z_2(t)$ in Table 1) is target output controllable and also functional stabilizable whilst it is not functional controllable.
	\item The functional $z_3(t)$ which constitute only controllable states  (see the row corresponding to $z_3(t)$ in Table 1) is target output controllable and functional stabilizable and also functional controllable.
	\item The functional $z_4(t)$ which is a mix of stabilizable states and uncontrollable states  (see the row corresponding to $z_4(t)$ in Table 1) is not target output controllable, not functional stabilizable and not functional controllable.
\end{itemize}

Although both \textit{target output controllability} and \textit{functional controllability} reduce to the standard notion of controllability when \( F = I_n \), the two concepts differ in their structural requirements.
{\it Functional controllability} requires that all functionals intended to be influenced must lie within a controllable subsystem, just as classical controllability requires that all states lie within the controllable subspace of the system.
In contrast, \textit{target output controllability} does \textit{not} impose this requirement. The functionals to be influenced do \textit{not} necessarily need to be part of the controllable subspace.
In this sense, \textit{functional controllability} is the concept that most naturally generalizes the classical notion of controllability.

The following lemma will be used in the sequel of the paper.
\begin{lemma}\label{lemma1}
	For matrices $X$ and $Y$ of appropriate dimensions, and $F$ full row rank, the following matrix $S$ where	
	\begin{equation} \label{eqS}
		S = \begin{pmatrix}
			X^{-} & I-X^{-}X &{\bf 0}\\
			Y &-YX &FF^T
		\end{pmatrix} 
	\end{equation}
	is of full row rank.
\end{lemma}
\begin{proof}
	Let the non-singular matrix $\Phi$ be
	\begin{equation}
		\Phi = \begin{pmatrix} I &X &{\bf 0} \\
			{\bf 0} &I &{\bf 0}\\
			-(FF^T)^{-1} Y &{\bf 0}  &(FF^T)^{-1} 
		\end{pmatrix}. \nonumber
	\end{equation} 
	Multiplying $S$ on the right by $\Phi$ yields
	\begin{IEEEeqnarray}{lll}
		\rank(S)	&&= \rank \begin{pmatrix}
			X^{-} & I-X^{-}X &{\bf 0}\\
			Y &-YX &FF^T
		\end{pmatrix} \Phi  \nonumber \\
		&&= \rank \begin{pmatrix}X^{-} &I &{\bf 0}\\
			{\bf 0} &{\bf 0} &I
		\end{pmatrix} \nonumber
	\end{IEEEeqnarray}	
	which proves the lemma.
\end{proof}

The following theorem highlights the distinction between target output controllability and functional controllability.
\begin{thm}
	If the triple \( (A,B,F^T) \) is functional controllable, then the triple \( (A,B,F) \) is target output controllable.
\end{thm}

\begin{proof}
	From Theorem 2, functional controllability of the triple $(A,B,F^T)$ is equivalent to
	\begin{equation}\label{11}
		\rank \begin{pmatrix}
			\mathcal{C}_{(A,B)} &F^T
		\end{pmatrix} = \rank(\mathcal{C}_{(A,B)})
	\end{equation}
	and we assume equation \eqref{11} is satisfied. 
This implies that the columns of \( F^T \) lie in the column space of \( \mathcal{C}_{(A,B)} \), so there exists a matrix \( \Omega \in \mathbb{R}^{r \times nm} \) with \( \rank(\Omega) = r \) such that
	\[
	F = \Omega \mathcal{C}_{(A,B)}^T.
	\]
	Then 
\begin{IEEEeqnarray}{rcl}
	&&\rank\begin{pmatrix}
		F\mathcal{C}_{(A,B)} &FF^T
		\end{pmatrix}
		 \nonumber\\
		&&= \rank\begin{pmatrix}
		\Omega\mathcal{C}_{(A,B)}^T\mathcal{C}_{(A,B)} &\Omega\mathcal{C}_{(A,B)}^T\mathcal{C}_{(A,B)}\Omega^T
	\end{pmatrix} \nonumber\\
    &&=\rank \Omega \mathcal{C}_{(A,B)}^T\mathcal{C}_{(A,B)}\begin{pmatrix}
    	I &\Omega^T
    \end{pmatrix}  =\rank\left(F\mathcal{C}_{(A,B)}\right). \label{eq4}
	\end{IEEEeqnarray} 
	Since $F$ is full row rank, $FF^T$ is invertible. According to Lemma \ref{lemma1}, $S$ in \eqref{eqS} is full row rank for $X=F\mathcal{C}_{(A,B)}$ and $Y=-(FF^T)^{-1}F\mathcal{C}_{(A,B)}\left(F\mathcal{C}_{(A,B)}\right)^-$, the left hand side of \eqref{eq4} can be rewritten as
	\begin{IEEEeqnarray}{rcl}
		&&\rank \begin{pmatrix}
			F\mathcal{C}_{(A,B)} &FF^T
		\end{pmatrix} 
		= \rank  \big(\begin{pmatrix}
			F\mathcal{C}_{(A,B)} &FF^T
		\end{pmatrix} S\big)\nonumber\\
		&&=\rank\begin{pmatrix}
			{\bf 0} &F\mathcal{C}_{(A,B)} &I_r
		\end{pmatrix} = \rank(F). \label{14}
	\end{IEEEeqnarray}
Item 2 of Theorem 1 follows directly from \eqref{eq4} and \eqref{14}, thereby completing the proof.
\end{proof}

The following theorem shows that functional controllability is a weaker condition than controllability.

\begin{thm}
	If the pair $(A,B)$ is controllable then the triple $(A,B,F^T)$ is functional controllable.
\end{thm}
\begin{proof}
	The pair $(A,B)$ being controllable is equivalent to	
	\begin{equation}
		\rank \mathcal{C}_{(A,B)} = n \nonumber
	\end{equation}	
	therefore
	\begin{equation}
		\rank\begin{pmatrix}\mathcal{C}_{(A, B)} &F^T\end{pmatrix} = \rank\big(\begin{array}{ll}\mathcal{C}_{(A, B)}\end{array}\big) = n. \nonumber
	\end{equation}
	This proves the theorem.
\end{proof}

The following theorem shows that functional stabilizability is a weaker condition than stabilizability.

\begin{thm}
	If the pair $(A,B)$ is stabilizable then the triple $(A,B,F^T)$ is functional stabilizable.
\end{thm}
\begin{proof}
	The pair $(A,B)$ being stabilizable is equivalent to		
	\begin{equation}
		\rank \left(\mathcal{C}_{(A,B)}\right) = n, \forall \lambda \in \C, \mathfrak{Re}(\lambda) \geq 0 \nonumber
	\end{equation}	
	therefore	
	\begin{IEEEeqnarray}{rcl}
		&&\rank\begin{pmatrix}(\lambda I-A)^n &\mathcal{C}_{(A, B)} &\mathcal{C}_{(A, F^T)}\end{pmatrix}\nonumber \\ 
		&&= \rank\big(\begin{array}{ll}\lambda I-A)^n &\mathcal{C}_{(A, B)}\end{array}\big)= n, \forall \lambda \in \C, \mathfrak{Re}(\lambda) \geq 0.\nonumber
	\end{IEEEeqnarray}
	This proves the theorem.	
\end{proof}

The remainder of this section presents a set of theorems that establish the duality of the concepts introduced earlier. It begins with formal definitions of functional observability and functional detectability, formulated analogously to Definitions 2, 3, 4, and 5. This is followed by the statement and proof of theorems related to functional observability and functional detectability.
\begin{definition}
		The functional \( z(t) = Fx(t) \) is observable, or the triple \( (A, C, F) \) is  \emph{functional observable}, if and only if every vector in the unobservable subspace of the pair \( (A,C) \) is also contained in the unobservable subspace of the pair \( (A,F) \). 
		That is,
		\[
		\ker(\mathcal{O}_{(A,C)}) \subseteq \ker(\mathcal{O}_{(A,F)}).
		\]
\end{definition}

\begin{definition}[Algebraic form of Definition 6]
The functional \( z(t) = Fx(t) \) is observable, or the triple \( (A, C, F) \) is \emph{functional observable}, if and only if for all \( v \in \mathbb{C}^n \)		
		\[
		\mathcal{O}_{(A,C)}v = \mathbf{0} \quad \Rightarrow \quad \mathcal{O}_{(A,F)}v  = \mathbf{0}
		\]
		where \( \mathcal{O}_{(A,C)} \) and \( \mathcal{O}_{(A,F)} \) denote the observability matrices corresponding to output matrix \( C \) and  matrix \( F, \) respectively.
\end{definition}

\begin{definition}
		The functional \( z(t) = Fx(t) \) is said to be detectable, or the triple \( (A, C, F) \) is \emph{functional detectable}, if and only if every generalized eigenvector of \( A \in \mathbb{R}^{n \times n} \) associated with an eigenvalue \( \lambda \in \mathbb{C} \) satisfying \( \mathfrak{Re}(\lambda) \geq 0 \), which lies in the unobservable subspace of the pair \( (A, C) \), also lies in the unobservable subspace of the pair \( (A, F) \).
\end{definition}

\begin{definition}[Algebraic form of Definition 8]
The linear combination of states \( z(t) = Fx(t) \) is detectable, or the triple \( (A, C, F) \) is \emph{functional detectable}, if and only if for all \( \lambda \in \mathbb{C} \) with \( \mathfrak{Re}(\lambda) \geq 0 \) and for all \( v \in \mathbb{C}^n  \)
		\begin{IEEEeqnarray}{rcl}
			&&(\lambda I - A)^nv  = \mathbf{0} \quad \text{and} \quad \mathcal{O}_{(A, C)}v  = \mathbf{0} \quad  \Rightarrow \quad \mathcal{O}_{(A, F)}v  = \mathbf{0}. \nonumber
		\end{IEEEeqnarray}
\end{definition}

Following theorem characterises Functional Observability.
\begin{thm}
	The triple \( (A, C, F) \) is functional observable if and only if \textnormal{(see \cite{ref12})}
	\[
	\rank\begin{pmatrix} \mathcal{O}_{(A,C)} \\ \mathcal{O}_{(A,F)} \end{pmatrix} = \rank\left( \mathcal{O}_{(A,C)} \right)
	\]
	or equivalently \textnormal{(see \cite{20a})}
	\[
	\rank\begin{pmatrix} \mathcal{O}_{(A,C)} \\ F \end{pmatrix} = \rank\left( \mathcal{O}_{(A,C)} \right).
	\]
\end{thm}

\begin{proof}
By Definition 6, the triple \( (A, C, F) \) is functional observable if and only if
		\[
		\ker\left( \mathcal{O}_{(A,C)} \right) \subseteq \ker\left( \mathcal{O}_{(A,F)} \right).
		\]
This can be restated as
		\[
		\ker\begin{pmatrix} \mathcal{O}_{(A,C)} \end{pmatrix} \subseteq \ker\begin{pmatrix} \mathcal{O}_{(A,C)} \\ \mathcal{O}_{(A,F)} \end{pmatrix}.
		\]
		Since appending more rows to a matrix can only reduce its kernel, the following inclusions also hold:
		\[
		\ker\begin{pmatrix} \mathcal{O}_{(A,C)} \\ \mathcal{O}_{(A,F)} \end{pmatrix} \subseteq \ker\begin{pmatrix} \mathcal{O}_{(A,C)} \\ F \end{pmatrix} \subseteq \ker\begin{pmatrix} \mathcal{O}_{(A,C)} \end{pmatrix}.
		\]
		Therefore, all three kernels must be equal:
		\[
		\ker\begin{pmatrix} \mathcal{O}_{(A,C)} \\ \mathcal{O}_{(A,F)} \end{pmatrix} = \ker\begin{pmatrix} \mathcal{O}_{(A,C)} \\ F \end{pmatrix} = \ker\left( \mathcal{O}_{(A,C)} \right).
		\]
		Given that the matrices have the same number of columns and equal kernels, the rank–nullity theorem guarantees that their ranks are equal, and vice versa:
		\[
		\rank\begin{pmatrix} \mathcal{O}_{(A,C)} \\ \mathcal{O}_{(A,F)} \end{pmatrix} = \rank\begin{pmatrix} \mathcal{O}_{(A,C)} \\ F \end{pmatrix} = \rank\left( \mathcal{O}_{(A,C)} \right)
		\]
		which proves the theorem.
\end{proof}

\begin{remark}[Historical Perspective]
The concept of \emph{functional observability} was introduced in~\cite{ref10} and further developed in~\cite{ref12}. The rank condition stated in Theorem 7 of the present paper was first reported in~\cite{ref12}  and then in \cite{ref11}, and more recently the equivalence of the two functional observability conditions in Theorem 7 was shown in \cite{ref13}. The proof of the rank condition in~\cite{ref12} (see Theorems 1 and 3 therein) is based on the unobservable subspace formulation, consistent with Definition 6 in this paper. On the other hand,  functional observability condition in Theorem 2 of~\cite{ref12} and Theorem 4 of \cite{ref11} is only valid for diagonalizable systems.
\end{remark}

The following theorem characterises functional detectability.
\begin{thm}
	The triple $(A, C, F)$	is functional detectable if and only if (see \cite{ref13})
	\begin{IEEEeqnarray}{rcl} 
		\rank \begin{pmatrix}
			(\lambda I -A)^n\\ \mathcal{O}_{(A,C)}\\ \mathcal{O}_{(A,F)}
		\end{pmatrix} &=& \rank \begin{pmatrix}
			(\lambda I -A)^n\\ \mathcal{O}_{(A,C)}
		\end{pmatrix}, \nonumber\\
		&& \forall \lambda \in \C, \mathfrak{Re}(\lambda) \geq 0 \nonumber
	\end{IEEEeqnarray}
	or equivalently 
	
	\begin{IEEEeqnarray}{rcl} 
		\rank \begin{pmatrix}
			(\lambda I -A)^n\\ \mathcal{O}_{(A,C)}\\ F
		\end{pmatrix} &=& \rank \begin{pmatrix}
			(\lambda I -A)^n\\ \mathcal{O}_{(A,C)}
		\end{pmatrix}, \nonumber\\
		&& \forall \lambda \in \C, \mathfrak{Re}(\lambda) \geq 0. \nonumber
	\end{IEEEeqnarray}
\end{thm}
\begin{proof}
By Definition 9, the triple \( (A, C, F) \) is functional detectable if and only if for each \( \lambda \in \mathbb{C} \) with \( \mathfrak{Re}(\lambda) \geq 0 \), 
		\[
		\ker \begin{pmatrix}
			(\lambda I -A)^n\\
			\mathcal{O}_{(A,C)}
		\end{pmatrix}
		\subseteq 
		\ker \begin{pmatrix}
			(\lambda I -A)^n\\
			\mathcal{O}_{(A,F)} 
		\end{pmatrix}.
		\]
This can be restated as
		\[
		\ker \begin{pmatrix}
			(\lambda I -A)^n\\
			\mathcal{O}_{(A,C)}
		\end{pmatrix}
		\subseteq 
		\ker \begin{pmatrix}
			(\lambda I -A)^n\\
			\mathcal{O}_{(A,C)}\\
			\mathcal{O}_{(A,F)} 
		\end{pmatrix}.
		\]
		Since appending more rows to a matrix cannot enlarge its kernel, the following inclusions also hold:
		\[
		\ker\begin{pmatrix} (\lambda I -A)^n\\\mathcal{O}_{(A,C)} \\ \mathcal{O}_{(A,F)} \end{pmatrix} \subseteq \ker\begin{pmatrix} (\lambda I -A)^n\\\mathcal{O}_{(A,C)} \\ F \end{pmatrix} \subseteq \ker\begin{pmatrix}(\lambda I -A)^n\\ \mathcal{O}_{(A,C)} \end{pmatrix}.
		\]
		Therefore, all three kernels must be equal:
		\[
		\ker\begin{pmatrix} (\lambda I -A)^n\\\mathcal{O}_{(A,C)} \\ \mathcal{O}_{(A,F)} \end{pmatrix} = \ker\begin{pmatrix} (\lambda I -A)^n\\\mathcal{O}_{(A,C)} \\ F \end{pmatrix} = \ker\begin{pmatrix} (\lambda I -A)^n\\\mathcal{O}_{(A,C)} \end{pmatrix}.
		\]
Given that the matrices have the same number of columns and equal kernels, the rank–nullity theorem guarantees that their ranks are equal, and vice versa:
		\begin{IEEEeqnarray}{rcl}
			\rank\begin{pmatrix} (\lambda I -A)^n\\\mathcal{O}_{(A,C)} \\ \mathcal{O}_{(A,F)} \end{pmatrix} &=& \rank\begin{pmatrix} (\lambda I -A)^n\\ \mathcal{O}_{(A,C)} \\ F \end{pmatrix}\nonumber\\
			&=& \rank\begin{pmatrix}
				(\lambda I -A)^n\\ \mathcal{O}_{(A,C)} \end{pmatrix}
		\end{IEEEeqnarray}
		which proves the theorem.
\end{proof}

The following theorem establishes the duality between functional controllability and functional observability.

\begin{thm}
	Functional controllability and functional observability are directly dual notions, with functional controllability of the system $(A, B, C)$ corresponding to functional observability of its dual system $(A^T, C^T, B^T)$, and vice versa.
\end{thm}
\begin{proof}
	The argument is presented in two parts. First, it is shown that if the triple \( (A, B, F^T) \) is functional controllable, then the triple \( (A^T, B^T, F) \) is functional observable. Second, it is demonstrated that if the triple \( (A, C, F) \) is functional observable, then the triple \( (A^T, C^T, F^T) \) is functional controllable.
	
	{\it (Part I)} - If the triple $(A, B, F^T)$ is functional controllable then
	\begin{IEEEeqnarray}{rcl}
		&&\rank\begin{pmatrix}\mathcal{C}_{(A,B)} &\mathcal{C}_{(A,F^T)}\end{pmatrix}=
		\rank \begin{pmatrix}\mathcal{C}_{(A,B)} \end{pmatrix}
		\nonumber
	\end{IEEEeqnarray}
	and can be rewritten as
	\begin{IEEEeqnarray}{rcl}
		&&\rank\begin{pmatrix}\mathcal{C}_{(A,B)} &\mathcal{C}_{(A,F^T)}\end{pmatrix}^T=
		\rank \begin{pmatrix}\mathcal{C}_{(A,B)} \end{pmatrix}^T
		\nonumber
	\end{IEEEeqnarray}
	and since
	\begin{IEEEeqnarray}{rcl}
		\rank\Big( \begin{pmatrix}\mathcal{C}_{(A,B)} &\mathcal{C}_{(A,F^T)}\end{pmatrix}^T\Big) &=& \rank \begin{pmatrix}
			B^T\\B^TA^T \\\vdots \\B^T(A^T)^{n-1} \\F \\FA^T\\\vdots \\F(A^T)^{n-1}
		\end{pmatrix} \nonumber\\
		&=&  \rank \begin{pmatrix}
			\mathcal{O}_{(A^T,B^T)}\\\mathcal{O}_{(A^T,F)}
		\end{pmatrix} \nonumber
	\end{IEEEeqnarray}
	and
	\begin{equation}\label{17}
		\rank \left(\mathcal{C}_{(A,B)}^T\right) = \rank \left(\mathcal{O}_{(A^T,B^T)}\right) \nonumber
	\end{equation}
	leads to 
	\begin{IEEEeqnarray}{rcl}
		\rank \begin{pmatrix}\mathcal{O}_{(A^T,B^T)}\\ \mathcal{O}_{(A^T,F)}\end{pmatrix} = \rank \begin{pmatrix}\mathcal{O}_{(A^T,B^T)}\end{pmatrix} \nonumber
	\end{IEEEeqnarray}
	which proves that if the triple \( (A, B, F^T) \) is functional controllable, then the triple \( (A^T, B^T, F) \) is functional observable.
	
	{\it (Part II)} - If the triple $(A, C, F)$ is functional observable then
	\begin{IEEEeqnarray}{rcl}
		\rank \begin{pmatrix}\mathcal{O}_{(A, C)}\\ \mathcal{O}_{(A, F)}\end{pmatrix} = \rank \begin{pmatrix}\mathcal{O}_{(A, C)}\end{pmatrix} \nonumber
	\end{IEEEeqnarray}
	and by taking the transpose of matrices on both sides, leads to 
	\begin{equation}
		\rank \begin{pmatrix}\mathcal{C}_{(A^T,C^T)} & \mathcal{C}_{(A^T,F^T)}\end{pmatrix}= \rank \left(\mathcal{C}_{(A^T,C^T)}\right)  \nonumber
	\end{equation}
	which proves if the triple \( (A, C, F) \) is functional observable, then the triple \( (A^T, C^T, F^T) \) is functional controllable.
\end{proof}	

The following theorem establishes the duality between functional stabilizability and functional detectability.

\begin{thm}
	Functional stabilizability and functional detectability are directly dual notions, with functional stabilizability of the system $(A, B, C)$ corresponding to functional detectability of its dual system $(A^T, C^T, B^T)$, and vice versa.
\end{thm}
\begin{proof}
	The argument is presented in two parts. First, it is shown that if the triple \( (A, B, F^T) \) is functional stabilizable, then the triple \( (A^T, B^T, F) \) is functional detectable. Second, it is demonstrated that if the triple \( (A, C, F) \) is functional detectable, then the triple \( (A^T, C^T, F^T) \) is functional stabilizable.
	
	{\it (Part I)} - From Theorem 3, functional stabilizability of system $(A, B, C)$ is equivalent to its Item 2, and taking the transpose of the matrix $\begin{pmatrix}
		(\lambda I -A)^n &\mathcal{C}_{(A,B)} &\mathcal{C}_{(A,F^T)}
	\end{pmatrix}$ which is on the left hand side of Item 2 of Theorem 3, and also taking the transpose of the matrix $\begin{pmatrix}
		(\lambda I -A)^n &\mathcal{C}_{(A,B)} 
	\end{pmatrix}$ which is on the right hand side of Item 2 of Theorem 3, leads to 
	\begin{IEEEeqnarray}{rcl}
		\rank \begin{pmatrix}
			(\lambda I -A^T)^n\\ \mathcal{O}_{(A^T,B^T)}\\ \mathcal{O}_{(A^T,F)}
		\end{pmatrix} &=& \rank \begin{pmatrix}
			(\lambda I -A^T)^n\\ \mathcal{O}_{(A^T,B^T)}
		\end{pmatrix}, \nonumber\\
		&& \forall \lambda \in \C, \mathfrak{Re}(\lambda) \geq 0
	\end{IEEEeqnarray}
	which is in fact the functional detectability (see Theorem 8) of the triple $(A^T, B^T, F)$.    
	
	{\it (Part II)} - From Theorem 8, functional detectability of the triple $(A,C,F)$ is equivalent to
	\begin{IEEEeqnarray}{rcl}\label{22}
		\rank \begin{pmatrix}
			(\lambda I -A)^n\\ \mathcal{O}_{(A,C)}\\ \mathcal{O}_{(A,F)}
		\end{pmatrix} &=& \rank \begin{pmatrix}
			(\lambda I -A)^n\\ \mathcal{O}_{(A,C)}
		\end{pmatrix}, \nonumber\\
		&& \forall \lambda \in \C, \mathfrak{Re}(\lambda) \geq 0
	\end{IEEEeqnarray}
	and taking the transpose of matrices on the left hand side and the right hand side of \eqref{22}, leads to 
	\begin{IEEEeqnarray}{rcl}\label{23}
		&&\rank \begin{pmatrix}
			(\lambda I -A^T)^n & \mathcal{C}_{(A^T,C^T)} & \mathcal{C}_{(A^T,F^T)}
		\end{pmatrix} = \nonumber\\
		&&\rank \begin{pmatrix}
			(\lambda I -A^T)^n & \mathcal{C}_{(A^T,C^T)}
		\end{pmatrix},  \forall \lambda \in \C, \mathfrak{Re}(\lambda) \geq 0
	\end{IEEEeqnarray}
	which is in fact the functional stabilizability (see Item 2 of Theorem 3) of the triple $(A^T, C^T, F^T)$.
\end{proof}

\begin{figure*}[!t]
	\centering
	\includegraphics[width=15.0cm, trim=30 70 40 40]{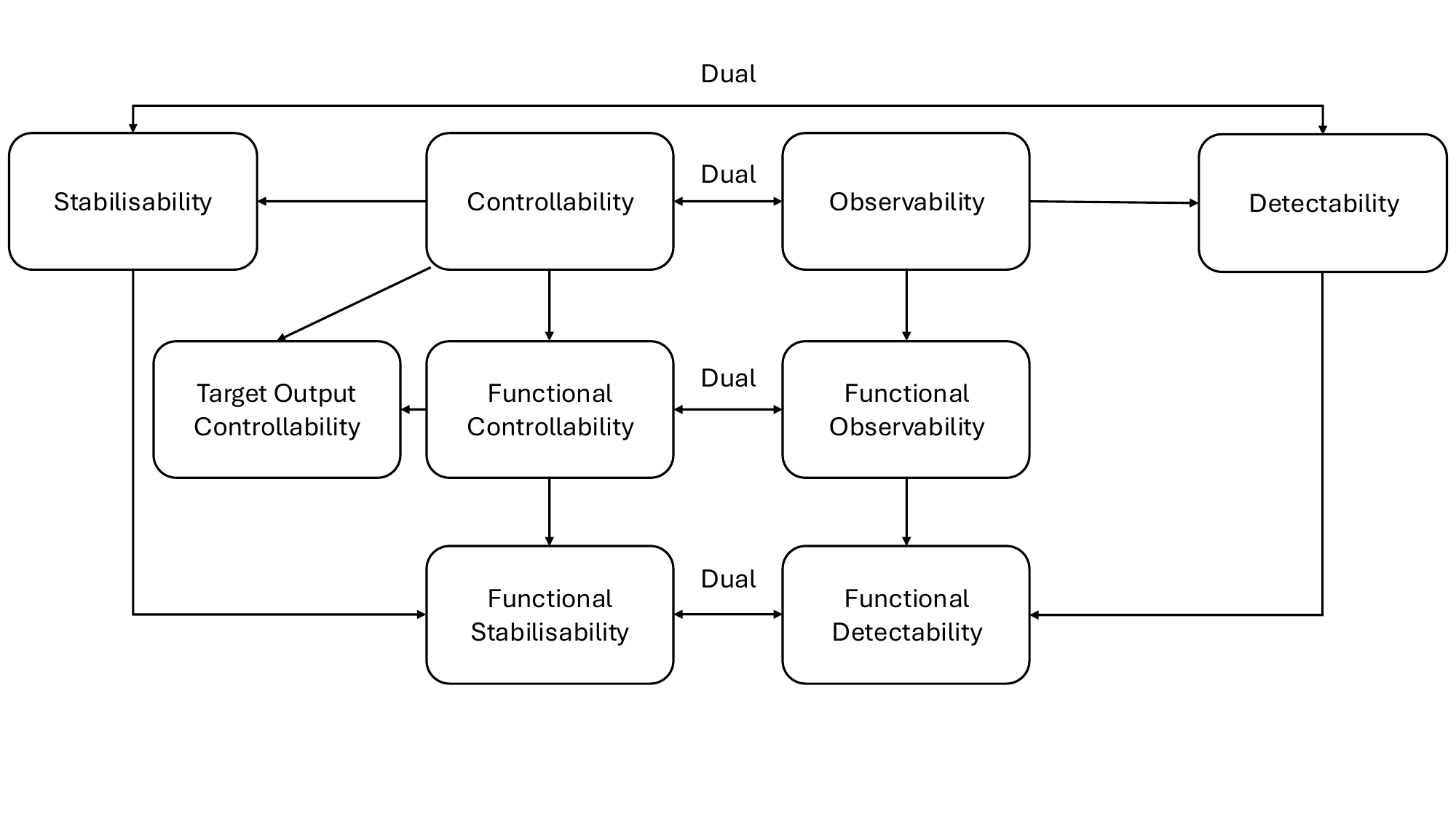} % Replace "diagram.pdf" with your PDF file name
	\centering
	\caption{Relationships among controllability, target output controllability, functional controllability, stabilizability, functional stabilizability, observability, functional observability, functional detectability, and detectability.}
	\label{fig1}
\end{figure*}

Based on the theorems and corollaries presented in this section, the commutative diagram in Figure \ref{fig1} illustrates the relationships among controllability, target output controllability, functional controllability, stabilizability, functional stabilizability, observability, functional observability, detectability and functional detectability. 

\section{Functional Controller Design by Pole Placement}
This section focuses on the design of functional controllers by pole placement. Pre-multiplying \eqref{1a} by $F$ yields
\begin{equation} 
	F\dot{x}(t) = FAx(t) + FBu(t) \nonumber
\end{equation}	
which can be rewritten as
\begin{IEEEeqnarray}{rcl}
	F\dot{x} &=& FA\left(I-F^-F+F^-F\right)x(t)+FBu(t) \nonumber\\
	&=& FAF^-Fx(t)+FA\left(I-F^-F\right)x(t)+FBu(t). \nonumber \\ \label{29}
\end{IEEEeqnarray}

The following two lemmas and the subsequent theorem were originally presented in \cite{1b} in the context of target output controller design. Since the result is equally applicable to functional controller design, it is restated here for completeness. The proof of the theorem is also included to preserve logical continuity in the development of the results that follow.

\begin{lemma}[\cite{1b}]
	The rank condition
	\[
	\rank\begin{pmatrix}
		FA\\F
	\end{pmatrix} = \rank(F)
	\]
	is equivalent to
	\begin{equation} \label{eq:14}
		FA(I - F^-F) = \mathbf{0}.
	\end{equation}
	If this condition holds, then the following identity is satisfied for all \( \lambda \in \mathbb{C} \):
	\begin{IEEEeqnarray}{rcl}
		\rank\begin{pmatrix}
			\lambda F - FA & FB
		\end{pmatrix} &=& 
		\rank\begin{pmatrix}
			\lambda I - FAF^- & FB
		\end{pmatrix}. \label{eq:15}
	\end{IEEEeqnarray}
\end{lemma}

\begin{lemma}[\cite{1b}]
	The following equation
	\begin{equation} \label{neweq52} 
		MF-FA=\mathbf{0}
	\end{equation} 
	where $A \in \R^{n\times n}$, $F\in \R^{r\times n}$, $\mathrm{rank}(F)=r$ and $r \leq n$ are known matrices and 
	$M \in \R^{r\times r}$ is an unknown matrix, has a solution
	if and only if
	
	\begin{equation} \label{neweqn54a}
		\mathrm{rank}\begin{pmatrix}
			FA\\F
		\end{pmatrix} =\mathrm{rank}(F)
	\end{equation}
	and in this situation $M$ satisfies
	\begin{equation}
		\mathrm{eig}(M) \subseteq \mathrm{eig}(A).
	\end{equation}
\end{lemma}	

\begin{thm}[\cite{1b}]
	Let \( Z \in \mathbb{R}^{m \times r} \) be a feedback gain matrix. The control law
	\[
	u(t) = -ZF x(t)
	\]
	drives the linear functional \( Fx(t) \to \mathbf{0} \) as \( t \to \infty \) with an arbitrary rate of convergence from any initial condition \( Fx(t_0) \), by assigning \( r \) eigenvalues to a subsystem of order \( r \), if and only if the following conditions hold:
	\begin{IEEEeqnarray}{rcl} \label{31}
		\mathrm{rank} \begin{pmatrix}
			FA \\
			F
		\end{pmatrix} &=& \mathrm{rank}(F) \IEEEyessubnumber \label{31a} \\
		\mathrm{rank} \begin{pmatrix}
			\lambda F - FA & FB
		\end{pmatrix} &=& \mathrm{rank}(F), \forall \lambda \in \mathbb{C}. \IEEEyessubnumber \label{31b}
	\end{IEEEeqnarray}
	Furthermore, the assigned eigenvalues form a subset of the eigenvalues of the closed-loop matrix \( A - BZF \).
\end{thm}

\begin{proof} {\it (Sufficiency)}
- If \eqref{31a} is satisfied or equivalently $FA(I-F^-F) = \mathbf{0}$ (see equation \eqref{eq:14} in Lemma 2) then
	equation \eqref{29} reduces to
	\begin{IEEEeqnarray}{rcl}
		F\dot{x} &=& FAF^-Fx(t)+FBu(t). \label{34}
	\end{IEEEeqnarray}
	If $u(t)=-ZFx(t)$, it now follows that $Fx(t) \rightarrow \mathbf{0}$ as $t \rightarrow \infty$ at an arbitrary convergence rate by placing $r$ poles if and only if the pair  $(FAF^-, FB)$ is controllable, which is equivalent to $\rank \begin{pmatrix}\lambda I-FAF^- &FB\end{pmatrix} = \rank(F) = r \,\,\forall\,\, \lambda \in \C$, which is in fact \eqref{31b} (see equation \eqref{eq:15} in Lemma 2). 
	
	{\it (Necessity)} - The contrapositive is employed to establish necessity. Specifically, if \eqref{31a} holds but \eqref{31b} does not—i.e., the pair \( (FAF^-, FB) \) is not controllable—then it follows that \( Fx(t) \not\rightarrow \mathbf{0} \) as \( t \rightarrow \infty \) at an arbitrary convergence rate.
	
	Moreover, under condition \eqref{31a}, $\left(FAF^{-}\right)F-FA=\mathbf{0}$ (see equation \eqref{eq:14} in Lemma 2), which can be rewritten as the following Sylvester equation
	\begin{equation}
		(FAF^{-} -FBZ)F-F(A-BZF)=\mathbf{0}. \label{neweqn55}
	\end{equation}
	Since $F \neq \mathbf{0}$ is full row rank, from Lemma 3 it follows that $\mathrm{eig}(FAF^{-} -FBZ) \subseteq \mathrm{eig}(A - BZF)$. This proves the theorem.
\end{proof}

\begin{thm}
	If Conditions \eqref{31a} and \eqref{31b} are satisfied, then the triple \( (A, B, F^T) \) is functional controllable, that is
		\[
		\ker\left(\mathcal{C}_{(A,B)}^T\right) \subseteq \ker\left(\mathcal{C}_{(A,F^T)}^T\right).
		\]
\end{thm}

\begin{proof}
		From Theorem~11, satisfaction of Conditions \eqref{31a} and \eqref{31b} implies that the control law \( u(t) = -ZFx(t) \), with \( Z \in \mathbb{R}^{m \times r} \), yields the closed-loop dynamics
		\[
		\dot{z}(t) = (FAF^- - FBZ)z(t), \quad z(t) = Fx(t).
		\]
		Since the pair \( (FAF^-, FB) \) is controllable (by \eqref{31b}), the eigenvalues of the matrix \( FAF^- - FBZ \) can be arbitrarily assigned. In particular, they can be placed in the open left-half complex plane so that
		\[
		\lim_{t \to \infty} Fx(t) = \mathbf{0}
		\]
		with an arbitrary rate of convergence for all initial states \( x(0) \in \mathbb{R}^n \).
	
		Now suppose \( v \in \ker\left(\mathcal{C}_{(A,B)}^T\right) \), i.e.,
		\[
		v^T A^k B = 0 \quad \text{for all } k = 0, \dots, n-1.
		\]
		Assume, for contradiction, that \( v \notin \ker\left(\mathcal{C}_{(A,F^T)}^T\right) \). Then there exists \( k \in \{0,\dots,n-1\} \) such that
		\[
		v^T A^k F^T \neq 0
		\]
		implying that \( v \) contributes to the functional \( z(t) = Fx(t) \), but cannot be influenced by the input through \( B \). This contradicts the fact that \( Fx(t) \to \mathbf{0} \) with an arbitrary rate of convergence for all initial conditions under the control law \( u(t) = -ZFx(t) \). Hence, the assumption must be false, and
		\[
		v \in \ker\left(\mathcal{C}_{(A,F^T)}^T\right).
		\]
		It follows that
		\[
		\ker\left(\mathcal{C}_{(A,B)}^T\right) \subseteq \ker\left(\mathcal{C}_{(A,F^T)}^T\right)
		\]
		which, by definition, implies that the triple \( (A, B, F^T) \) is functional controllable.
\end{proof}

\section{Generalized Separation Principle}
The Separation Principle is a fundamental result in control and estimation theory, stating that state estimation and control design can be treated independently in linear time-invariant systems. When all system states must be controlled and estimated, the Separation Principle ensures that the controller and state estimator can be designed separately if and only if the system is both controllable and observable.

This section presents a generalization of the Separation Principle by extending it to the estimation and control of specific desired functionals, even when the full system is neither controllable nor observable. The necessary and sufficient condition for independently estimating and controlling such functionals is shown to be related to functional observability and functional controllability. Moreover, when the selected functionals coincide with the entire state vector, the generalized Separation Principle reduces to the classical Separation Principle.

Recall that the Darouach observer of order \( r \), used for estimating the desired functionals \( z(t) \), takes the following form (see \cite{22new}):
\begin{IEEEeqnarray}{rcl}
	\dot{w}(t) &=& N_rw(t)+J_ry(t)+H_ru(t) \IEEEyessubnumber \label{eq:87a}\\
	\hat{z}(t) &=& w(t) + E_ry(t)	\IEEEyessubnumber \label{eq:87b}
\end{IEEEeqnarray}
where $w(t)\in \mathbb{R}^r$, $\hat{z}(t)\in \mathbb{R}^r$ is the estimate of $z(t)$, matrices $N_r$, $J_r$, $H_r$ and $E_r$ have appropriate dimensions. The existence of the functional observer of order $r$ is given in the following theorem:
\begin{thm}[\cite{22new}]
	The necessary and sufficient conditions for the existence of a Darouach observer of order \( r \), such that the estimation error converges to zero at an arbitrarily prescribed rate, are given by:
	\begin{IEEEeqnarray}{rcl}\label{87}
		&&\rank \begin{pmatrix}
			FA\\
			CA\\
			C\\
			F
		\end{pmatrix} = 	\rank \begin{pmatrix}
			CA\\
			C\\
			F
		\end{pmatrix} \IEEEyessubnumber \label{87a}\\
		&&\rank \begin{pmatrix}
			\lambda F - FA\\
			CA\\
			C
		\end{pmatrix} = 	\rank \begin{pmatrix}
			CA\\
			C\\
			F
		\end{pmatrix}, \forall \lambda \in \mathbb{C}. \IEEEyessubnumber \label{87b}
	\end{IEEEeqnarray}
\end{thm}

A notable feature of the Darouach observer structure is that its order matches the number of functionals to be estimated, corresponding to the number of rows in the matrix $F$. Furthermore, when conditions \eqref{87a} and \eqref{87b} are satisfied, the estimation error \( e(t) = z(t) - \hat{z}(t) \) evolves according to (see equations (3)-(5) in \cite{22new}):
\begin{equation}\label{92}
	\dot{e}(t) = N_r e(t)
\end{equation}
where the eigenvalues of the matrix \( N_r \in \mathbb{R}^{r \times r} \) can be arbitrarily assigned in the complex plane. Once conditions \eqref{87a} and \eqref{87b} are met, the observer parameters \( N_r \), \( J_r \), \( H_r \), and \( E_r \) can be readily computed, as outlined in \cite{22new}.

The conditions \eqref{87a} and \eqref{87b}, which are necessary and sufficient for designing a functional observer of order 
$r$ (by placing $r$ observer poles), are analogous to conditions \eqref{31a} and \eqref{31b}, which ensure the existence of a functional controller capable of assigning $r$ closed-loop poles, as stated in Theorem~11. Together, these conditions represent parallel theoretical results: the former pertains to the existence of a Darouach-type functional observer of order $r$, while the latter pertains to the existence of a functional controller achieving placement of $r$ closed-loop poles.

%The conditions \eqref{87a} and \eqref{87b}, which are necessary and sufficient for designing a functional observer of order \( r \) (involving the placement of \( r \) observer poles), are analogous to the conditions \eqref{31a} and \eqref{31b}, which are necessary and sufficient to design a functional controller that assigns \( r \) closed-loop poles, as presented in Theorem~9. These conditions thus constitute parallel theoretical results: the former set ensures the existence of the Darouach functional observer which is of order $r$, while the latter guarantees the existence of a functional controller that places $r$ closed loop poles.

The situation is now addressed in which the conditions \eqref{87a} and \eqref{87b} required for functional observer design, or the conditions \eqref{31a} and \eqref{31b} required for functional controller design, are not satisfied.
First, consider the case where conditions \eqref{87a} and \eqref{87b} are not met. The observer design can be relaxed using an {\it augmentation approach} \cite{ref10}, which builds on the observer structure \eqref{eq:87a} and \eqref{eq:87b}. In this approach, the matrix \( F \) is augmented with a freely chosen matrix \( R \), and the observer design is carried out using the augmented matrix:
\[
\begin{pmatrix} F \\ R \end{pmatrix}
\]
which is used in place of \( F \) in conditions \eqref{87a} and \eqref{87b}. Consequently, the observer is designed to be of order \( q \), corresponding to the number of rows in the augmented matrix \( \begin{pmatrix} F \\ R \end{pmatrix} \). The observer then estimates the extended functional:
\begin{equation}\label{eq84}
	\begin{pmatrix} z(t) \\ z_R(t) \end{pmatrix} = \begin{pmatrix} F \\ R \end{pmatrix} x(t)
\end{equation}
and it follows from \eqref{eq:87a} and \eqref{eq:87b} that the observer of order $q$ takes the following form \cite{ref10}:
\begin{IEEEeqnarray}{rcl}
	\dot{w}(t) &=& N_qw(t)+J_qy(t)+H_qu(t) \IEEEyessubnumber \label{91a}\\
	\begin{pmatrix} \hat{z}(t) \\ \hat{z}_R(t) \end{pmatrix} &=&w(t) + E_qy(t)	\IEEEyessubnumber \label{91b}
\end{IEEEeqnarray}
where $w(t)\in \mathbb{R}^q$, $\hat{z}(t)\in \mathbb{R}^r$ is the estimate of $z(t)$, $\hat{z}_R(t)\in \mathbb{R}^{q-r}$ is the estimate of $z_R(t)$, matrices $N_q$, $J_q$, $H_q$ and $E_q$ have appropriate dimensions. By incorporating the arbitrary matrix \( R \) into the extended functional \eqref{eq84}, the observer of order \( q \) is effectively used to estimate the original functional \( z(t) \) through the structure given in \eqref{91a}-\eqref{91b}. Moreover, the estimation error, $\epsilon(t)$, of the observer of order \( q \) is given by 
\begin{equation}
	\epsilon(t) = \begin{pmatrix} z(t) \\ z_R(t) \end{pmatrix} - \begin{pmatrix} \hat{z}(t) \\\hat{z}_R(t) \end{pmatrix}.
\end{equation}

The following corollary follows directly from Theorem 13.

\begin{cor}[\cite{ref10}]
	A functional observer of order \(q\) exists for the estimation of the extended functional
	\[
	\begin{pmatrix} z(t) \\ z_R(t) \end{pmatrix} = \begin{pmatrix} F \\ R \end{pmatrix} x(t),
	\]
	such that the estimation error converges to zero at an arbitrarily prescribed rate, if and only if there exists a matrix \( R \in \mathbb{R}^{q-r} \), with \( q \in \{r, \dots, n\} \), satisfying the following two conditions:
	\begin{IEEEeqnarray}{rcl} \label{eq:34}
		& \rank \begin{pmatrix}
			FA \\
			RA \\
			CA \\
			C \\
			F \\
			R
		\end{pmatrix} = \rank \begin{pmatrix}
			CA \\
			C \\
			F \\
			R
		\end{pmatrix} \IEEEyessubnumber \label{92a} \\[1em]
		& \rank \begin{pmatrix}
			\lambda F - FA \\
			\lambda R - RA \\
			CA \\
			C
		\end{pmatrix} = \rank \begin{pmatrix}
			CA \\
			C \\
			F \\
			R
		\end{pmatrix}, \quad \forall \lambda \in \mathbb{C} \IEEEyessubnumber \label{92b}
	\end{IEEEeqnarray}
	
	Additionally, the estimation error, given by \( \epsilon(t) \), evolves according to
	\[
	\dot{\epsilon}(t) = N_q \epsilon(t),
	\]
	where the eigenvalues of the matrix \( N_q \in \mathbb{R}^{q \times q} \) can be freely assigned anywhere in the complex plane.
\end{cor}
\begin{remark}
	Conditions \eqref{92a} and \eqref{92b} are more relaxed than \eqref{87a} and \eqref{87b}, offering flexibility in choosing \(R\), which allows for an observer of order \(q \geq r\) to estimate $z(t)$.
\end{remark}

	The following theorem formalizes the conditions under which such a functional observer exists.
	
	\begin{thm}
		The triple \( (A, C, F) \) is functional observable if and only if there exists \( q \in \mathbb{N} \) such that \( r \leq q \leq n \), and a matrix \( R \in \mathbb{R}^{(q - r) \times n} \) (which is empty if \( q = r \) and non-empty if \( r < q \leq n \)), such that conditions \eqref{92a} and \eqref{92b} are satisfied.
	\end{thm}

\begin{proof}
{\it (Sufficiency)} - 
		Assume there exists an integer \( q, \, r \leq q \leq n \), and a matrix \( R \in \mathbb{R}^{(q - r) \times n} \) such that the extended functional matrix \( \bar{F} = \begin{pmatrix} F \\ R \end{pmatrix} \in \mathbb{R}^{q \times n} \) satisfies conditions \eqref{92a} and \eqref{92b}. By Corollary 4, these rank conditions guarantee the existence of a functional observer of order \( q \) that estimates \( \bar{F}x(t) \) asymptotically. Since \( Fx(t) \) is a subvector of \( \bar{F}x(t) \), it follows that \( Fx(t) \) is also asymptotically reconstructible from \( y(t) = Cx(t) \), implying that \( Fx(t) \) lies within the observable subspace of the pair \( (A, C) \). This is equivalent to:
		\[
		\operatorname{Im}(F^T) \subseteq \operatorname{Im}(\mathcal{O}_{(A,C)}^T)
		\]
		or, equivalently:
		\[
		\rank \begin{pmatrix} \mathcal{O}_{(A,C)} \\ F \end{pmatrix} = \rank \left( \mathcal{O}_{(A,C)} \right),
		\]
		which is the functional observability condition. Hence, the triple \( (A, C, F) \) is functional observable.
	
	\medskip
	
{\it (Necessity)} - 
		Assume \( (A, C, F) \) is functional observable. Then, there exists a similarity transformation bringing the system into the Kalman observability decomposition:
		\[
		T^{-1} A T = \begin{pmatrix} A_o & \mathbf{0} \\ A_{21} & A_u \end{pmatrix}, \quad C T = \begin{pmatrix} C_o & \mathbf{0} \end{pmatrix},
		\]
		where \( A_o \in \mathbb{R}^{q \times q} \) governs the observable subspace and \( A_u \) the unobservable subspace. Since \( Fx(t) \) lies within the observable subspace, the matrix \( F \) takes the form:
		\[
		F T = \begin{pmatrix} F_o & \mathbf{0} \end{pmatrix}, \quad F_o \in \mathbb{R}^{r \times q}.
		\]
		Define \( R \in \mathbb{R}^{(q - r) \times n} \) as
		\[
		R = \begin{pmatrix} F^\perp_o & \mathbf{0} \end{pmatrix} T^{-1}
		\]
		and
		\[
		\bar{F}_o = \begin{pmatrix} F_o \\ F_o^\perp \end{pmatrix}.
		\]
		Then, the extended functional matrix becomes
		\[
		\bar{F} = \begin{pmatrix} F \\ R \end{pmatrix} = \begin{pmatrix} F_o & \mathbf{0} \\ F_o^\perp & \mathbf{0} \end{pmatrix} T^{-1} = \begin{pmatrix} \bar{F}_o & \mathbf{0} \end{pmatrix} T^{-1}.
		\]
		The rank conditions \eqref{92a} and \eqref{92b} can now be verified:
		\begin{itemize}
			\item For condition \eqref{92a}, observe:
			\[
			\rank \begin{pmatrix} \bar{F}A \\ \bar{F} \\ CA \\ C \end{pmatrix} = q \quad \text{and} \quad \rank \begin{pmatrix} CA \\ C \\ \bar{F} \end{pmatrix} = q.
			\]
			Hence, \eqref{92a} holds.		
			\item For condition \eqref{92b}, for all \( \lambda \in \mathbb{C} \), observe:
			\[
			\rank \begin{pmatrix} \lambda \bar{F} - \bar{F}A \\ CA \\ C \end{pmatrix} = q \quad \text{and} \quad \rank \begin{pmatrix} CA \\ C \\ \bar{F} \end{pmatrix} = q.
			\]
			Thus, \eqref{92b} holds.
		\end{itemize}
This completes the proof.
\end{proof}

\begin{remark}{\it (Consistency with prior definition of functional observability)} -- 
		Theorem~14 is consistent with the original definition of functional observability introduced in~\cite{ref10} and~\cite{ref12}, where the property is characterised by the existence of  \( q \in \mathbb{N},\,\,r \leq q \leq n \) and a matrix \( R \in \mathbb{R}^{(q - r) \times n} \), such that rank conditions \eqref{92a} and \eqref{92b} are satisfied. The existence of such an \( R \) is, in fact, equivalent to the functional observability condition presented in Theorem~7, as established in Theorem~14. In Theorem~7, this condition is derived from a geometric perspective, based on the inclusion of unobservable subspaces associated with the pairs \( (A, C) \) and \( (A, F) \), namely
		\[
		\operatorname{Im}(\mathcal{O}_{(A,F)}^T) \subseteq \operatorname{Im}(\mathcal{O}_{(A,C)}^T).
		\]
		This subspace-based formulation provides a transparent link to classical observability theory, while the construction involving \( R \) offers a constructive method for verifying or designing functional observers of a prescribed order. Both formulations are equivalent and reflect the same fundamental requirement: that the functional \( Fx(t) \) be asymptotically reconstructible from the measured output \( y(t) = Cx(t) \).
\end{remark}
	\begin{remark}
		The necessity proof of Theorem~14 shows that if the triple \( (A, C, F) \) is functional observable, then a matrix \( R \) satisfying conditions \eqref{92a} and \eqref{92b} {\it can always be constructed}. One such choice is
		\[
		R = \begin{pmatrix} F^\perp_o & \mathbf{0} \end{pmatrix} T^{-1}
		\]
		where \( T \) is any similarity transformation that brings the system into Kalman observable form. This construction is not unique; alternatives with lower row dimension for $R$ may also satisfy the same conditions. A method for searching such alternatives for $R$ is reported in \cite{ref10}.
	\end{remark}

Similarly, if conditions \eqref{31a} and \eqref{31b} are not satisfied, the requirements for controlling the linear functional \( Fx(t) \) by placing \( r \) poles can be relaxed by augmenting matrix \( F \) with an additional matrix \( R_1 \in \R^{(q_1-r)\times n}\), $q_1 > r$. That is, instead of using \( F \), it is replaced with the augmented matrix:
\[
\begin{pmatrix} F \\ R_1 \end{pmatrix}
\]
in \eqref{31a} and \eqref{31b} as per the following corollary derived from Theorem 11.

\begin{cor}
	The control law $$u(t)=-Z\begin{pmatrix}F\\R_1\end{pmatrix}x(t) = -Z \begin{pmatrix}\hat{z}(t)\\\hat{z}_{R_1}(t)\end{pmatrix}$$
	where $Z\in \R^{m\times q_1}$ can drive the 
	linearly independent functions $\begin{pmatrix}\hat{z}(t)\\\hat{z}_{R_1}(t)\end{pmatrix} \rightarrow \mathbf{0}$ as $t \rightarrow \infty$ at an arbitrary rate of convergence from any initial condition $\begin{pmatrix}\hat{z}(t_0)\\\hat{z}_{R_1}(t_0)\end{pmatrix}$ by placing $q_1$ poles of a subsystem of order $q_1$ if and only if the following conditions are satisfied:
	\begin{IEEEeqnarray}{rcl}\label{eq:35}
		\rank\begin{pmatrix} 	
			FA\\R_1A\\F\\R_1
		\end{pmatrix} &=& \rank\begin{pmatrix}F\\R_1\end{pmatrix} \IEEEyessubnumber \label{94a}\\
		\rank\begin{pmatrix}
			\lambda F-FA &FB\\\lambda R_1-R_1A &R_1B
		\end{pmatrix} &=& \rank\begin{pmatrix}F\\R_1\end{pmatrix}, \forall \lambda \in \C. \IEEEyessubnumber \label{94b}
	\end{IEEEeqnarray} 
	Furthermore, the assigned eigenvalues form a subset of the eigenvalues of the closed-loop matrix \( A - BZ\begin{pmatrix}
		F\\R_1
	\end{pmatrix} \).
\end{cor}
\begin{proof}
	Replace $F$ with $\begin{pmatrix}F\\R_1\end{pmatrix}$ and $r$ with $q_1$ in the proof of Theorem 11.
\end{proof}	

Now let
\begin{equation}
	R=\begin{pmatrix}
		R_1\\R_2
	\end{pmatrix}.
\end{equation}

\begin{remark}
	When \( R = \emptyset \), conditions \eqref{92a}, \eqref{92b}, \eqref{94a}, and \eqref{94b} reduce to \eqref{87a}, \eqref{87b}, \eqref{31a}, and \eqref{31b}, respectively. Hence, conditions \eqref{92a}, \eqref{92b}, \eqref{94a}, and \eqref{94b} can be regarded as generalizations of the conditions \eqref{87a}, \eqref{87b}, \eqref{31a}, and \eqref{31b}.
\end{remark}

The Generalized Separation Principle is now presented in the following theorem:

\begin{thm}
	Let \( R_1 \) and \( R_2 \) be matrices that is either empty or nonempty. If there exists such \( R_1 \) and \( R_2 \) for which the conditions \eqref{eq:34} and \eqref{eq:35} are satisfied, then a functional observer and a functional controller can be independently designed such that \( z(t) \rightarrow \mathbf{0} \) at an arbitrarily prescribed rate of convergence as \( t \rightarrow \infty \).
\end{thm}

\begin{proof}
	Pre-multiplying \eqref{1a} by $ \begin{pmatrix} F \\ R_1 \end{pmatrix}$, leads to
	\begin{equation} 
		\begin{pmatrix} F \\ R_1 \end{pmatrix}\dot{x}(t) = \begin{pmatrix} F \\ R_1 \end{pmatrix}Ax(t) +  \begin{pmatrix} F \\ R_1 \end{pmatrix}Bu(t) \nonumber 
	\end{equation}	
	which can be written as
	\begin{IEEEeqnarray}{rcl}
		&&\begin{pmatrix} F \\ R_1 \end{pmatrix}\dot{x} = \begin{pmatrix} F \\ R_1 \end{pmatrix}Bu(t) \nonumber\\
		&&+  \begin{pmatrix} F \\ R_1 \end{pmatrix}A\left(I- \begin{pmatrix} F \\ R_1 \end{pmatrix}^- \begin{pmatrix} F \\ R_1 \end{pmatrix}+ \begin{pmatrix} F \\ R_1 \end{pmatrix}^- \begin{pmatrix} F \\ R_1 \end{pmatrix}\right)x(t)\nonumber\\
		&=&  \begin{pmatrix} F \\ R_1 \end{pmatrix}A \begin{pmatrix} F \\ R_1 \end{pmatrix}^- \begin{pmatrix} F \\ R_1 \end{pmatrix}x(t) + \begin{pmatrix} F \\ R_1 \end{pmatrix}Bu(t)\nonumber\\
		&&+ \begin{pmatrix} F \\ R_1 \end{pmatrix}A\left(I- \begin{pmatrix} F \\ R_1 \end{pmatrix}^- \begin{pmatrix} F \\ R_1 \end{pmatrix}\right)x(t). 
		\label{eq96}
	\end{IEEEeqnarray}	
	If \eqref{94a} is satisfied or equivalently $ \begin{pmatrix} F \\ R_1 \end{pmatrix}A\left(I- \begin{pmatrix} F \\ R_1 \end{pmatrix}^- \begin{pmatrix} F \\ R_1 \end{pmatrix}\right) = \mathbf{0}$ (see equation \eqref{eq:14} in Lemma 2) then
	equation \eqref{eq96} reduces to
	\begin{IEEEeqnarray}{rcl}
		\begin{pmatrix} F \\ R_1 \end{pmatrix}\dot{x} &=&  \begin{pmatrix} F \\ R_1 \end{pmatrix}A \begin{pmatrix} F \\ R_1 \end{pmatrix}^- \begin{pmatrix} F \\ R_1 \end{pmatrix}x(t)+ \begin{pmatrix} F \\ R_1 \end{pmatrix}Bu(t). \nonumber\\ \label{97}
	\end{IEEEeqnarray}
	Since $
	\begin{pmatrix} z(t) \\ z_{R_1}(t) \end{pmatrix} = \begin{pmatrix} F \\ R_1 \end{pmatrix} x(t),
	$ equation \eqref{97} can be written as follows:
	
	\begin{IEEEeqnarray}{rcl}
		\begin{pmatrix} \dot{z}(t) \\ \dot{z}_{R_1}(t) \end{pmatrix} &=&  \begin{pmatrix} F \\ R_1 \end{pmatrix}A \begin{pmatrix} F \\ R_1 \end{pmatrix}^- \begin{pmatrix} z(t) \\ z_{R_1}(t) \end{pmatrix}+ \begin{pmatrix} F \\ R_1 \end{pmatrix}Bu(t). \nonumber \\
		\label{98}
	\end{IEEEeqnarray}	
	In an observer based controller design, $u(t)$ is chosen as follows:
	\begin{IEEEeqnarray}{rcl}
		u(t) &=& -Z\begin{pmatrix}\hat{z}(t)\\\hat{z}_{R_1}(t)\end{pmatrix} =-Z\left(\begin{pmatrix}z(t)\\z_{R_1}(t)\end{pmatrix} - \begin{pmatrix}
			I_{q_1} &\bf{0}
		\end{pmatrix}\epsilon(t)\right) \nonumber
	\end{IEEEeqnarray}
	and substituting it in \eqref{98} leads to
	\begin{IEEEeqnarray}{rcl}\label{103}
		\begin{pmatrix} \dot{z}(t) \\ \dot{z}_{R_1}(t) \end{pmatrix} &=& \left( \begin{pmatrix} F \\ R_1 \end{pmatrix}A \begin{pmatrix} F \\ R_1 \end{pmatrix}^- - \begin{pmatrix} F \\ R_1 \end{pmatrix}BZ\right)\begin{pmatrix} z(t) \\ z_{R_1}(t) \end{pmatrix} \nonumber\\
		&&+\begin{pmatrix} F \\ R_1 \end{pmatrix}BZ \begin{pmatrix}
				I_{q_1} &\bf{0}
		\end{pmatrix}\epsilon(t)  
	\end{IEEEeqnarray}	
	Combining the observer-based control system \eqref{103} with the observer error dynamics, \( \dot{\epsilon}(t) = N_q \epsilon(t) \), yields the following expression:
	\begin{equation}\label{eqn104}
		\begin{pmatrix} 
			\dot{z}(t) \\ \dot{z}_{R_1}(t)\\\dot{\epsilon}(t) \end{pmatrix} = 
		\Psi\left(\begin{array}{c} z(t) \\ z_{R_1}(t)\\\hdashline\epsilon(t) \end{array}\right)
	\end{equation}
	where 	
	\begin{equation}
		\Psi = \left( \begin{array}{c:c} \begin{pmatrix}F \\ R_1 \end{pmatrix}A \begin{pmatrix} F \\ R_1 \end{pmatrix}^--\begin{pmatrix} F \\ R_1 \end{pmatrix}BZ &\begin{pmatrix} F \\ R_1 \end{pmatrix}BZ\begin{pmatrix}
					I_{q_1} &\bf{0}
			\end{pmatrix}\\\hdashline
			\bf{0} &N_q	
		\end{array}\right). \nonumber 
	\end{equation}
	If conditions \eqref{31a}, \eqref{31b}, \eqref{87a} and \eqref{87b} are satisfied then choosing $R=\emptyset$ ensures conditions \eqref{92a}, \eqref{92b}, \eqref{94a} and \eqref{94b} are also satisfied. In this scenario, when $R=\emptyset$, the combined observer based control system and the observer error dynamics in \eqref{eqn104} reduces to
	\begin{equation}\label{eqn106}
		\begin{pmatrix} 
			\dot{z}(t) \\ \dot{e}(t) \end{pmatrix} = 
		\left( \begin{array}{c:c} FA F^--FBZ &FBZ\\\hdashline
			\bf{0} &N_r	
		\end{array}\right)\left(\begin{array}{c} z(t) \\e(t) \end{array}\right).
	\end{equation}
	This closed loop system matrix in \eqref{eqn104} and \eqref{eqn106} are upper triangular, and therefore the eigenvalues of the full system are the union of the eigenvalues of \( \left(\begin{pmatrix}F \\ R_1 \end{pmatrix}A \begin{pmatrix} F \\ R_1 \end{pmatrix}^- - \begin{pmatrix} F \\ R_1 \end{pmatrix}BZ \right)\) and \( N_q \). Hence, the system is asymptotically stable if and only if both \( \left(\begin{pmatrix}F \\ R_1 \end{pmatrix}A \begin{pmatrix} F \\ R_1 \end{pmatrix}^- - \begin{pmatrix} F \\ R_1 \end{pmatrix}BZ \right)\) and \( N_q \) are stable which requires the satisfaction of conditions \eqref{eq:34} and \eqref{eq:35} for some $R$ either empty or nonempty. Since the eigenvalues of \( \left(\begin{pmatrix}F \\ R_1 \end{pmatrix}A \begin{pmatrix} F \\ R_1 \end{pmatrix}^- - \begin{pmatrix} F \\ R_1 \end{pmatrix}BZ \right)\) and \( N_q \)  are chosen independently, the controller and observer can be designed separately. This proves the generalized separation principle.
\end{proof} 
	
If \( F = I \), then \( z(t) = x(t) \). Moreover, even when \( R = \emptyset \), the left- and right-hand sides of condition \eqref{92a} reduce to:

\begin{equation} \label{102}
	\rank \begin{pmatrix}
		FA\\
		CA\\
		C\\
		F
	\end{pmatrix} = 	\rank \begin{pmatrix}
		CA\\
		C\\
		F
	\end{pmatrix} = n.
\end{equation}
Similarly, the left-hand side of condition \eqref{92b} becomes:
\begin{IEEEeqnarray}{rcl}
	\rank \begin{pmatrix}
		\lambda F - FA\\
		CA\\
		C
	\end{pmatrix} &=& \rank \left(
	\begin{pmatrix}
		I & \bf{0} & \bf{0} \\
		C & I & -\lambda I \\
		\bf{0} & \bf{0} & I
	\end{pmatrix}
	\begin{pmatrix}
		\lambda I - A\\
		CA\\
		C
	\end{pmatrix}
	\right) \nonumber \\
	&=& \rank \begin{pmatrix}
		\lambda I - A\\
		C
	\end{pmatrix}.
\end{IEEEeqnarray}
Since \( \rank \begin{pmatrix} CA\\ C\\ F \end{pmatrix} = n \), condition \eqref{92b} becomes:

\begin{equation} \label{104}
	\rank \begin{pmatrix}
		\lambda I - A\\
		C
	\end{pmatrix} =n, \forall \lambda \in \mathbb{C}.
\end{equation}
Furthermore, condition \eqref{94a} becomes:

\begin{equation}
	\rank \begin{pmatrix}
		FA\\
		F
	\end{pmatrix} = \rank(F) = n
\end{equation}
and condition \eqref{94b} simplifies to:

\begin{equation} \label{106}
	\rank \begin{pmatrix}
		\lambda I - A & B
	\end{pmatrix} = n, \forall \lambda \in \mathbb{C}.
\end{equation}

\begin{remark}
	From equations \eqref{102}-\eqref{106}, it follows that for \( F = I \) and \( R = \emptyset \), the satisfaction of conditions \eqref{eq:34} and \eqref{eq:35} reduces to the classical observability condition \eqref{104} and controllability condition \eqref{106}. Therefore, when \( F = I \), Theorem 15 reduces to the well-known separation principle.
\end{remark}

If conditions~\eqref{31} and~\eqref{87} are satisfied, then the choice \( R = \emptyset \) satisfies both~\eqref{eq:34} and~\eqref{eq:35}. However, when~\eqref{31a} and~\eqref{87a} are not satisfied, it is necessary to determine a non-empty matrix 
\[
R = \begin{pmatrix} R_1 \\ R_2 \end{pmatrix}
\] 
\noindent such that the rank conditions~\eqref{92a} and~\eqref{94a} are satisfied. In particular, if~\eqref{31a} fails, then a valid choice for \( R_1 \) can be constructed based on the observability indices of the pair \( (A, F) \).

Let \( \nu_1, \dots, \nu_r \) denote the observability indices of the pair $(A,F)$ associated with the rows \( F_1, \dots, F_r \) of the matrix
\[
F = \begin{pmatrix}
	F_1\\
	\vdots\\
	F_r
\end{pmatrix}.
\]

\begin{thm}
Let \( R = \begin{pmatrix} R_1 \\ R_2 \end{pmatrix} \) be a matrix that satisfies condition~\eqref{92a}, and let \( R_1 \) satisfy condition~\eqref{94a}. Then one valid choice for \( R_1 \) and \( R_2 \) is given by
		\begin{IEEEeqnarray}{rcl}\label{ty39}
			R_1 &=& \Lambda = 
			\begin{pmatrix}
				\Lambda_1 \\
				\vdots \\
				\Lambda_r
			\end{pmatrix} \IEEEyessubnumber \label{eq:47} \\
			R_2 &=& \emptyset \IEEEyessubnumber
	\end{IEEEeqnarray}
where, for each \( i = 1, \dots, r \), the block \( \Lambda_i \in \mathbb{R}^{(\nu_i - 1) \times n} \) is defined as
	\[
	\Lambda_i = 
	\begin{pmatrix}
		F_i A \\
		F_i A^2 \\
		\vdots \\
		F_i A^{\nu_i - 1}
	\end{pmatrix}.
	\]
\end{thm}

\begin{proof}
	By construction,
	
	\[
	\rank\begin{pmatrix}
		F\\R
	\end{pmatrix} = \sum_{i=1}^{r} 
	\rank\begin{pmatrix}
		F_i\\\Lambda_i
	\end{pmatrix} = \sum_{i=1}^{r} \nu_i
	\]
	since \( \nu_i \) is the observability index of \( F_i \) with respect to \( A \). Thus, for each \( i \),
	\[
	\rank\begin{pmatrix}
		F_i A^{\nu_i} \\
		F \\
		\Lambda
	\end{pmatrix}
	= 
	\rank\begin{pmatrix}
		F \\
		\Lambda
	\end{pmatrix}.
	\]
	Stacking these vertically yields
	\[
	\rank\begin{pmatrix}
		F_1 A^{\nu_1} \\
		\vdots \\
		F_r A^{\nu_r} \\
		F \\
		\Lambda
	\end{pmatrix}
	= 
	\rank\begin{pmatrix}
		F \\
		\Lambda
	\end{pmatrix}
	= 
	\rank\begin{pmatrix}
		F \\
		R
	\end{pmatrix}
	= \sum_{i=1}^{r} \nu_i
	\]
	which confirms that condition~\eqref{94a} is satisfied. Moreover, any matrix \( R_1 \) satisfying~\eqref{94a} also satisfies~\eqref{92a} when \( R = R_1 \), completing the proof.
\end{proof}

\begin{remark}
The construction above provides one valid choice for \( R_1 \) and \( R_2 \), but the solution is not unique. Other matrices, including those with \( R_2 \neq \emptyset \), can also satisfy the required rank conditions in~\eqref{92a} and~\eqref{94a}.
\end{remark}

			\begin{remark}
					In \cite{3a}, the control law $u(t) = -Zx(t)$, which utilises the entire state space, is proposed to achieve target control of $z(t) = Fx(t)$. This is different from the control signal proposed in this paper:
					\[
					u(t) = -Z \begin{pmatrix} F \\ R_1 \end{pmatrix} x(t)
					\]
					where $R_1$ is selected to satisfy the conditions \eqref{92a}, \eqref{92b}, \eqref{94a}, and \eqref{94b}. However, if rows of $R_1$ forms a basis for the orthogonal complement of $F$ (i.e., 
					$F^\perp$) in which case $R_2 = \emptyset$, and if $R$ satisfies conditions \eqref{92a}, \eqref{92b}, \eqref{94a}, and \eqref{94b}
					then the proposed feedback controller also utilises the entire state space, aligning with the structure of the control signal in \cite{3a}.
			\end{remark}
			
			\begin{remark}
					According to Theorem~14, functional observability of the triple \( (A, C, F) \) is equivalent to the existence of a matrix \( R \) such that the rank conditions~\eqref{92a} and~\eqref{92b} are satisfied.
					However, functional controllability of the triple \( (A, B, F^T) \) does not necessarily guarantee the existence of a matrix \( R_1 \) such conditions~\eqref{94a} and~\eqref{94b} are satisfied. This can be demonstrated using a simple numerical example. Let			
					\[
					A = \begin{pmatrix}
						1 & 1 & 1 \\
						0 & 2 & 1 \\
						0 & 0 & 1
					\end{pmatrix}, \quad
					B = \begin{pmatrix}
						1 \\
						2 \\
						0
					\end{pmatrix}, \quad
					F = \begin{pmatrix}
						1 & 1 & 0
					\end{pmatrix}.
					\]
					The triple \( (A, B, F^T) \) is functional controllable. In order to satisfy~\eqref{94a}, the matrix \( R_1 \) must include rows constructed as in~\eqref{eq:47}; for this example, alternative solutions for \( R_1 \) also include any matrix such that
					
					\[
					\rank \begin{pmatrix} F \\ R_1 \end{pmatrix} = 3.
					\]
					However, with		
					
					\[
					R_1 = \begin{pmatrix}
						1 & 3 & 2 \\
						1 & 7 & 6
					\end{pmatrix}
					\]
					or with any other such \( R_1 \), it is straightforward to verify that condition~\eqref{94b} is not satisfied. This illustrates that, even though the triple \( (A, B, F^T) \) is functional controllable, it is not always possible to find a matrix \( R_1 \) that simultaneously satisfies both~\eqref{94a} and~\eqref{94b}.
			\end{remark}

			While this work lays a foundational framework for the design of functional controllers and functional observers -- independent of classical notions of controllability and observability -- it also introduces new theoretical challenges. In particular, although the matrix $R$ as per Theorem 16 satisfies conditions \eqref{92a} and \eqref{94a}, its ability to satisfy conditions \eqref{92b} and \eqref{94b} is not guaranteed. The systematic construction of an augmentation matrix $R$ with a minimum number of rows that ensures all four conditions -- \eqref{92a}, \eqref{92b}, \eqref{94a}, and \eqref{94b} -- are satisfied remains an open problem. This is identified as a direction for future research.

			\section{Numerical Examples}
			\noindent {\it Example 1 continued: illustration of the duality between functional controllability/functional stabilizability and functional observability/functional detectability.} 
			
			Example 1 from Section III is continued to illustrate the duality between functional controllability and functional observability, as well as the corresponding duality between functional stabilizability and functional detectability. Let		
			
			$$C=B^T=\begin{pmatrix}
				1 &0 &0&0\\
				0 &1&0 &0
			\end{pmatrix}
			$$
			$$F=\begin{pmatrix}
				1 &1 &0&0
			\end{pmatrix}.
			$$
			From Theorem 2, the triple $(A, B, F^T)$ is functional controllable, and by Theorem 7, the dual triple $(A^T, B^T, F)$ is functional observable. Similarly, Theorem 7 establishes that $(A, C, F)$ is functional observable, while Theorem 2 confirms that $(A^T, C^T, F^T)$ is functional controllable. These results, as summarised in Theorem 9, demonstrate the duality between functional controllability and functional observability.
			
			To illustrate the duality between functional stabilizability and functional detectability, consider the following matrix $F$:			
			$$F=\begin{pmatrix}
				1 &0 &1&0
			\end{pmatrix}.
			$$
			From Theorem 3, the triple $(A, B, F^T)$ is functional stabilizable, and by Theorem 8, the dual triple $(A^T, B^T, F)$ is functional detectable. Similarly, Theorem 8 establishes that $(A, C, F)$ is functional detectable, while Theorem 3 confirms that $(A^T, C^T, F^T)$ is functional stabilizable. These results, as summarised in Theorem 10, demonstrate the duality between functional stabilizability and functional detectability.\\
			
			\noindent {\it Example 2: illustration of the generalized separation principle for the case $R = \emptyset$.} 
			
			Consider the following system, which is both uncontrollable and unobservable:\\
			\\
			$
			A = \begin{pmatrix}
				0.25 & 2.25 & 0.75 & -0.25 & 1.50 \\
				2.25 & 0.25 & -0.25 & 0.75 & -1.50 \\
				1.75 & 1.75 & 0.25 & 1.25 & -0.50 \\
				-1.25 & -1.25 & 2.25 & 1.25 & 0.50 \\
				0 & 0 & 0 & 0 & -4.00
			\end{pmatrix}
			$\\
			\\
			
			$
			B=\left( \begin{array}{rr}
				2\\
				0\\
				0\\
				0\\
				0 
			\end{array}\right)
			$, $C = \begin{pmatrix}1 & 1 & 0 & 0 & 0 \end{pmatrix}$ and let
			$$F = \begin{pmatrix}0.5 &0.5 &0.5 &0.5 &0\end{pmatrix}.$$ 
			\noindent {\it (Functional Controller Design for the case where $R=\emptyset$):}
			
			Theorem~2 is first used to establish that the triple \( (A, B, F^T) \) is functionally controllable. Condition~\eqref{31a} is then verified, implying that condition~\eqref{94a} holds with \( R_1 = \emptyset \).
			Moreover, the pair $(FAF^-, FB)$ is controllable or equivalently, condition \eqref{94b} is satisfied for $R_1=\emptyset$. Using pole placement, the value \( Z = 6 \) is selected to assign the eigenvalues of \( FAF^- - FBZ \) at \( -3 \).
			A functional controller of the form $u(t)=-FBZz(t)$ is utilised according to Corollary 5 and substituting it in \eqref{98}, leads to		
			
			$$\dot{z}(t)=(FAF^--FBZ)z(t)=-3z(t).$$
			It is observed that the eigenvalue $-3$, assigned through the control design, now appears in the spectrum of the closed-loop system matrix $A - BZF$. The original eigenvalues of the open-loop system matrix $A$ are $-4, -1, -2, 2, 3$. Under the control law $u(t) = -FBZz(t) = -FBZF x(t)$, the eigenvalues of the closed-loop matrix $A - BZF$ become $-4, -3, -2, -1, 2$.\\
			
			\noindent {\it (Functional Observer Design for the case where $R=\emptyset$):} 
			
			By Theorem~7, the triple \( (A, C, F) \) is functionally observable, and condition \eqref{92a} is satisfied with \( R = \emptyset \). Moreover, condition \eqref{92b} also holds for \( R = \emptyset \). Choosing the observer pole at \( -6 \) (see \cite{22new}) yields the following observer parameters: 
		
			\[
			N_r = -6, \quad E_r = 9, \quad K_r = -18, \quad J_r = -72, \quad H_r = -17.
			\]
			The corresponding functional observer error equation, according to \eqref{92}, is 
			\[
			\dot{e}(t) = -6e(t).
			\]
			By \eqref{eqn106}, the observer-based control system and the observer error dynamics can be combined as follows:		
			
			\begin{equation}
				\begin{pmatrix} 
					\dot{z}(t) \\ \dot{e}(t) \end{pmatrix} = 
				\left( \begin{array}{cc} -3&6\\
					0 &-6
				\end{array}\right)\left(\begin{array}{c} z(t) \\ e(t) \end{array}\right). \nonumber
			\end{equation}
			\\
			\noindent {\it Example 3:  illustration of the generalized separation principle for the case $R \neq \emptyset$.} 

			Consider the same uncontrollable and unobservable system presented in Example~2, with the following matrix \( F \):
			
			$$F = \begin{pmatrix}1.5 &1.5 &-0.5 &-0.5 &0\end{pmatrix}.$$ 
			
			\noindent{\it (Functional Controller Design for the case where $R\neq\emptyset$):}\\
			By Theorem~2, the triple \( (A, B, F^T) \) is established to be functionally controllable. However, condition~\eqref{31a} is not satisfied, implying that \( R \neq \emptyset \). Applying Theorem~16, the following is determined:		
			
			$$R=R_1=FA=\begin{pmatrix}
				3.5 &3.5 &-0.5 &-0.5 &0
			\end{pmatrix}$$
			and verify that condition \eqref{94a} is satisfied. Moreover, the pair $\left(\begin{pmatrix}
				F\\R_1
			\end{pmatrix}A\begin{pmatrix}
				F\\R_1
			\end{pmatrix}^-, \begin{pmatrix}
				F\\R_1
			\end{pmatrix}B\right)$ is controllable or equivalently, condition \eqref{94b} is satisfied. Using pole placement, the matrix		
		\[
	Z = \begin{pmatrix} -148.5 & 65.5 \end{pmatrix}
\]
is selected to place the eigenvalues of

\[
\left( \begin{pmatrix} F \\ R_1 \end{pmatrix} A \begin{pmatrix} F \\ R_1 \end{pmatrix}^- - \begin{pmatrix} F \\ R_1 \end{pmatrix} B Z \right)
\]
at \( -3 \) and \( -5 \). A functional controller of the form \( u(t) = -\begin{pmatrix} F \\ R_1 \end{pmatrix} B Z z(t) \) is employed in accordance with Corollary~5. Substituting this into \eqref{98} yields:

			\begin{IEEEeqnarray}{rcl}
			\dot{z}(t) &=&\left(\begin{pmatrix}
				F\\R_1
			\end{pmatrix}A\begin{pmatrix}
			F\\R_1
		\end{pmatrix}^--\begin{pmatrix}
		F\\R_1
	\end{pmatrix}BZ\right)z(t) \nonumber\\
 &=&\begin{pmatrix}
				445.5 &-195.5\\1033.5 &-453.5
			\end{pmatrix}z(t).\nonumber
		\end{IEEEeqnarray}
			The assigned eigenvalues $-3$ and $-5$ have been successfully incorporated into the closed-loop system matrix $A - BZ\begin{pmatrix}
				F\\R_1
			\end{pmatrix}$. The original eigenvalues of the open-loop system matrix $A$ are $-4, -1, -2, 2$, and $3$. Under the control law \( u(t) = -\begin{pmatrix}
			F\\R_1
		\end{pmatrix}BZz(t) \), the eigenvalues of the closed-loop matrix \( A - BZ\begin{pmatrix}
		F\\R_1
	\end{pmatrix} \) become \( -5, -4, -3, -2, -1 \). This demonstrates that a functional controller has been designed to ensure closed-loop stability by assigning all eigenvalues to the open left-half of the complex plane. \\
			
			\noindent {\it (Functional Observer Design for the case where $R\neq\emptyset$)}:
			
			Using Theorem~7, it can be shown that the triple \( (A, C, L) \) is functional observable. Furthermore, one can verify that conditions \eqref{92a} and \eqref{92b} are satisfied with the choice \( R = FA \) as per equation \eqref{ty39} in Theorem 16. Based on this, and selecting the observer poles at \( -6 \) and \( -7 \), the corresponding observer parameters can be determined as described in \cite{22new}:\\ 
			\\
			$N_q=\begin{pmatrix}
				-6 &0\\0 &-7
			\end{pmatrix}, E_q=\begin{pmatrix}
				-7\\-6
			\end{pmatrix}, K_q=\begin{pmatrix}
				30\\48
			\end{pmatrix}, J_q=\begin{pmatrix}
				72\\90
			\end{pmatrix}$ and $H_q=\begin{pmatrix}
				17\\19
			\end{pmatrix}$.\\

			\noindent The functional observer error equation is				
			
			$$
			\dot{e}(t)=\begin{pmatrix}
				-6 &0\\0 &-7
			\end{pmatrix}e(t).
			$$
			Using \eqref{eqn106}, the observer-based control system and the observer error dynamics can be combined to yield the following expression:						
			\begin{equation}
				\begin{pmatrix} 
					\dot{z}(t) \\ \dot{e}(t) \end{pmatrix} = 
				\left( \begin{array}{rr:rr} 445.5&-195.5 &-445.5 &196.5\\
					1033.5 &-453.5 &-1039.5 &458.5\\ \hdashline
					0 &0 &-6 &0\\0 &0 &0 &-7
				\end{array}\right)\left(\begin{array}{c} z(t) \\ e(t) \end{array}\right) \nonumber
			\end{equation}
		
			\section{Conclusion}
This paper established the theoretical foundations of functional controllability and stabilizability, extending classical control theory to settings where only specific functionals of the state are of interest. Building on prior results in functional observability and detectability, it formalised the duality between these properties and introduced a Generalized Separation Principle, enabling the independent design of functional controllers and observers—even when the system is neither fully controllable nor observable. Necessary and sufficient conditions were derived for the existence of functional controllers and observers of a given order, along with corresponding design methods and illustrative examples. These results generalise the classical separation principle, which appears as a special case when the functionals span the full state vector. A key open problem remains: the systematic construction of an augmentation matrix \( R \) with minimal row dimension that ensures all four conditions—\eqref{92a}, \eqref{92b}, \eqref{94a}, and \eqref{94b}—are simultaneously satisfied. 

%		
%			
%			\section{Conclusion}
%			This paper developed the theoretical foundations of functional controllability and functional stabilisability, extending classical control theory to settings where only specific functionals of the state are of interest. Building on previously established results on functional observability and functional detectability, the duality between these functional concepts was established, leading to the formulation of a Generalized Separation Principle that enables the independent design of functional controllers and observers---even when the system is neither fully controllable nor observable.
%			Necessary and sufficient conditions were derived for the existence of functional controllers and observers of a given order, accompanied by corresponding design methodologies and illustrative examples. These results generalise the classical separation principle, which appears as a special case when the functionals span the full state vector.
%			A key open problem remains: the systematic construction of an appropriate augmentation matrix $R$ with the least number of rows that ensures all four conditions---\eqref{92a}, \eqref{92b}, \eqref{94a}, and \eqref{94b}---are simultaneously satisfied. Resolving this challenge would broaden the applicability of the proposed theoretical framework.

\newpage
\section*{Biography}
\begin{wrapfigure}{l}{1in}
	\includegraphics[width=1in,height=1.25in,clip,keepaspectratio]{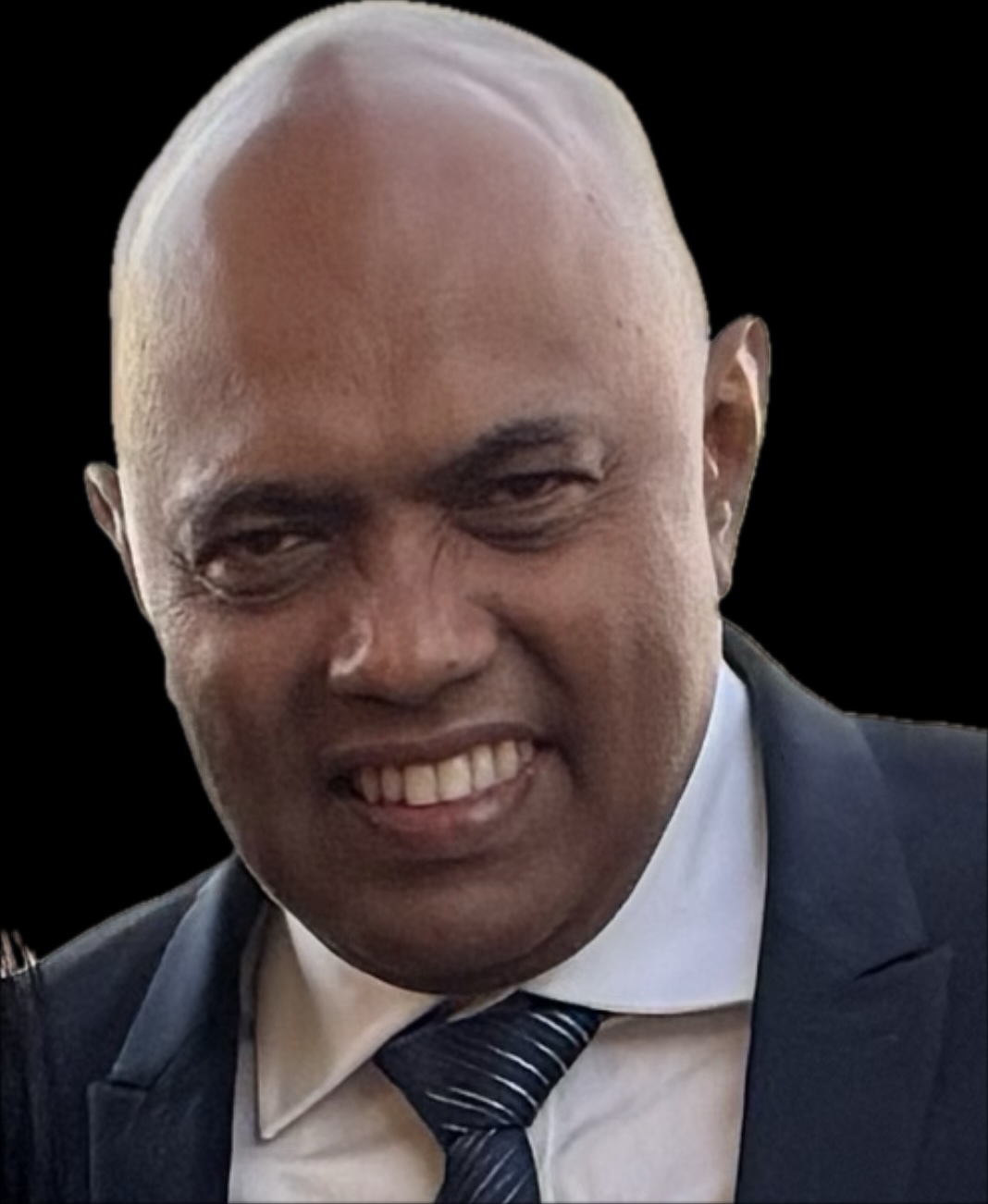}
\end{wrapfigure}
\noindent
\textbf{Tyrone Fernando} received his B.E. (Hons.) and Ph.D. degrees in Electrical Engineering from the University of Melbourne, Victoria, Australia, in 1990 and 1996, respectively. In 1996, he joined the Department of Electrical, Electronic and Computer Engineering at the University of Western Australia (UWA), Crawley, WA, Australia, where he currently holds the position of Professor of Electrical Engineering. He previously served as Associate Head and Deputy Head of the department from 2008 to 2010.

Prof. Fernando is currently the Head of the Power and Clean Energy Research Group at UWA. He has provided professional consultancy to Western Power on the integration and management of distributed energy resources in the electric grid. In recognition of his professional contributions, he was named the Outstanding WA IEEE PES/PELS Engineer in 2018. His research interests include theoretical control, observer design, and power system stability and control. He has received multiple teaching awards from UWA in recognition of his contributions to control systems and power systems education.
%He has also served as an Associate Editor for IEEE Transactions on Circuits and Systems I: Regular Papers and IEEE Transactions on Circuits and Systems II: Express Briefs. From 2019 to 2021, he was the Chair of the Power and Energy Technical Committee of the IEEE Circuits and Systems Society.

\begin{wrapfigure}{l}{1in}
	\includegraphics[width=1in,height=1.25in,clip,keepaspectratio]{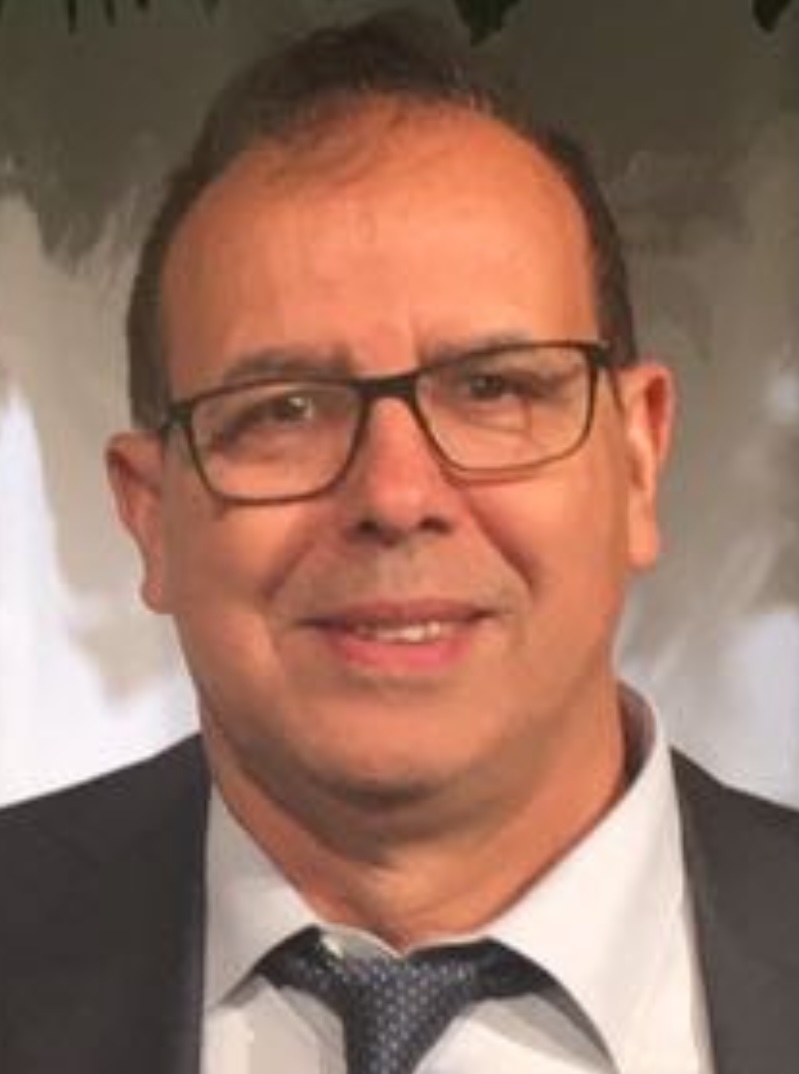}
\end{wrapfigure}

\noindent
\textbf{Mohamed Darouach} received his graduate degree in Electrical Engineering from Ecole Mohammadia d'Ingénieurs, Rabat, Morocco, in 1978. He earned the Doctor Engineer degree in Automatic Control and the Doctor of Sciences degree in Physics from Nancy University, Nancy, France, in 1983 and 1986, respectively. He was also awarded an honorary doctorate (Honoris Causa) from the Technical University of IASI, Iasi, Romania.

From 1978 to 1986, he served as Associate Professor and later Professor of Automatic Control at Ecole Hassania des Travaux Publics, Casablanca, Morocco. From 1987 to 2024, he was a Professor at Université de Lorraine, France, and has held the title of Distinguished Professor Emeritus at the same institution since 2024. Prof. Darouach served as Vice Director of the Research Center for Automatic Control of Nancy (CRAN UMR 7039, Université de Lorraine, CNRS) from 2005 to 2013. Since 2010, he has been a member of the Scientific Council at the University of Luxembourg. He also served as Vice Director of the University Institute of Technology of Longwy (Université de Lorraine) from 2013 to 2018.
He has held visiting positions at the University of Alberta, Edmonton, Canada, and the University of Western Australia (UWA), Perth, Australia. Since 2017, he has been a Robert and Maude Gledden Senior Visiting Fellow, administered by the Institute of Advanced Studies at UWA. He was also a Forrest Visiting Fellow at UWA in 2019, 2022, and 2023. His research interests include theoretical control, observer design, and the control of large-scale systems with applications.

\section*{References}
\vspace{-0.6cm}
\begin{thebibliography}{99} 
\bibitem{ref1}
R.~E. Kalman, ``On the general theory of control systems,'' in 
\emph{Proceedings of the First International Conference on Automatic Control}, Moscow, USSR, 1960.
\bibitem{ref2}
R.~E. Kalman, ``Mathematical description of linear dynamical systems,''
\emph{Journal of the Society for Industrial and Applied Mathematics, Series A: Control},
vol.~1, no.~2, pp.~152--192, 1963.
\bibitem{ref3}
T.~Kailath, \emph{Linear Systems}.  
Englewood Cliffs, NJ: Prentice-Hall, 1980.
\bibitem{ref4} 
A.~K.~Singh and B.~C.~Pal, ``Decentralized dynamic state estimation in power systems using unscented transformation,'' \emph{IEEE Transactions on Power Systems}, vol.~29, no.~2, pp.~794--804, 2013.
\bibitem{ref5} 
A.~N.~Montanari, C.~Duan, L.~A.~Aguirre, and A.~E.~Motter, ``Functional observability and target state estimation in large-scale networks,'' \emph{Proceedings of the National Academy of Sciences of the United States of America}, vol.~119, no.~1, e2113750119, Jan. 2022. doi: 10.1073/pnas.2113750119.
\bibitem{ref6} 
H.~Xu, Z.~Chen, C.~Zhou, S.~Shen, and F.~Gao, ``Omni-Swarm: A decentralized omnidirectional visual--inertial--UWB state estimation system for aerial swarms,'' \emph{IEEE Transactions on Robotics}, vol.~38, no.~6, pp.~3374--3394, Dec.~2022, doi: 10.1109/TRO.2022.3182503.
\bibitem{16new}
Y.~Zhang, R.~Cheng, and Y.~Xia, ``On the structural output controllability and functional observability of undirected networks,'' \emph{Automatica}, vol.~173, Art.~no.~112063, 2025.
\bibitem{17new} Y.~Zhang, R.~Cheng, and Y.~Xia, ``Observability blocking for functional privacy of linear dynamic networks,'' \emph{arXiv:2304.07928}, 2023. 
\bibitem{15new}J.~Gao, Y.-Y.~Liu, R.~M.~D'Souza, and A.-L.~Barabási, ``Target control of complex networks,'' \emph{Nature Communications}, vol.~5, 2014, Art.~no.~5415.
\bibitem{ref10} T. Fernando, H. Trinh, L. Jennings, ``Functional observability and the design of minimum order linear functional observers'', {\it IEEE Trans. Autom. Contr.}, 55 (5), pp. 1268-1273, 2010.
\bibitem{ref12} 
T.~Fernando, L.~Jennings, and H.~Trinh, ``Functional Observability,'' in \emph{Proceedings of the 2010 Fifth International Conference on Information and Automation for Sustainability (ICIAfS)}, Colombo, Sri Lanka, 2010, pp.~421--423, doi:~10.1109/ICIAFS.2010.5715662.
\bibitem{ref11} L. Jennings, T. Fernando, H. Trinh, ``Existence conditions for functional observability from an eigenspace perspective'', {\it IEEE Trans. Autom. Contr.}, 56 (12), pp. 2957-2961, 2011.
\bibitem{20a} F. Rotella and I. Zambettakis, ``A note on functional observability", {\it IEEE Trans. Autom. Contr.}, 61 (10), pp. 3197-3202, 2016.
\bibitem{ref13} M. Darouach and T. Fernando, ``On functional observability and functional observer design",  {\it Automatica}, vol. 173, 2025, Art. no. 112115.
\bibitem{22new}  M. Darouach, ``Existence and design of functional observers'', {\it IEEE Trans. Autom. Contr.}, 45(5), pp. 940-943, 2000.
\bibitem{13new}
M.~Darouach and T.~Fernando, ``On the existence and design of functional observers,'' \emph{IEEE Transactions on Automatic Control}, vol.~65, no.~6, pp.~2751--2759, Jun. 2020.
\bibitem{rotella2011minimal}
F.~Rotella and I.~Zambettakis, ``Minimal single linear functional observers for linear systems,'' \emph{Automatica}, vol.~47, no.~1, pp.~164--169, 2011.
\bibitem{rotella2016direct}
F.~Rotella and I.~Zambettakis, ``A direct design procedure for linear state functional observers,'' \emph{Automatica}, vol.~70, pp.~211--216, 2016.
\bibitem{mohajerpoor2015reduced}
R.~Mohajerpoor, H.~Abdi, and S.~Nahavandi, ``Reduced-order functional observers with application to partial state estimation of linear systems with input-delays,'' \emph{Journal of Control and Decision}, vol.~2, no.~4, pp. 233--256, 2015.
\bibitem{23a} H. Trinh, ``Linear functional state observer for time-delay systems'', {\it International Journal of Control}, 72 (18), pp. 1642-1658, 1999.
\bibitem{19a} D.C. Huong, V.T. Huynh and H. Trinh, ``Interval Functional Observers Design for Time-Delay Systems Under Stealthy Attacks'', {\it IEEE Transactions on Circuits and Systems I: Regular Papers}, 67 (12), pp. 5101 -5112, 2020. 
\bibitem{3} H.~Trinh and T.~Fernando, \emph{Functional Observers for Dynamical Systems}, Lecture Notes in Control and Information Sciences. Springer, 2012.
\bibitem{1a} J. Bertram and P. Sarachik, ``On optimal computer control'', {\it IFAC Proceedings Volumes 1}, pp. 429-432, 1960.
\bibitem{10a}  M. Lazar and J. Loh\'{e}ac, ``Output controllability in a long-time
horizon'', {\it Automatica}, 113, 108762, 2020.
\bibitem{6a} B. Danhane, J. Loh\'{e}ac and M. Jungers, ``Characterizations of
output controllability for LTI systems'', {\it Automatica}, 154, 111104, 2023.
\bibitem{4a} A. N. Montanari, C. Duan, A. E. Motter, ``Target controllability and target observability of structured network systems'', {\it IEEE Control Systems Letters}, 7, pp. 3060-3065, 2023.
\bibitem{1b} T. Fernando and M. Darouach, ``Existence and design of target output controllers'', {\it IEEE Trans. Autom. Contr.}, 2025, doi: 10.1109/TAC.2025.3552031.
\bibitem{3a} A. N. Montanari, C. Duan, A. E. Motter, ``Duality between controllability and observability for target control and estimation in networks'', {\it IEEE Trans.~Autom.~Control}, 2025, doi: 10.1109/TAC.2025.3552001.
%\bibitem{ref24}
%David A.~Harville, \emph{Matrix Algebra from a Statistician's Perspective},  Springer Verlag, 2000.

%%%%%%%%%%%%%%%%%%%%%%%%%
\end{thebibliography}
\end{document}